\PassOptionsToPackage{unicode}{hyperref}
\PassOptionsToPackage{hyphens}{url}
\documentclass[twocolumn,10pt,a4paper,]{article}
\usepackage{xcolor}
\usepackage[margin=0.68in]{geometry}
\usepackage{amsmath,amssymb}
\usepackage{iftex}
\ifPDFTeX
  \usepackage[T1]{fontenc}
  \usepackage[utf8]{inputenc}
  \usepackage{textcomp} 
\else 
  \usepackage{unicode-math} 
  \defaultfontfeatures{Scale=MatchLowercase}
  \defaultfontfeatures[\rmfamily]{Ligatures=TeX,Scale=1}
\fi
\usepackage{lmodern}
\ifPDFTeX\else
\fi
\IfFileExists{upquote.sty}{\usepackage{upquote}}{}
\IfFileExists{microtype.sty}{
  \usepackage[]{microtype}
  \UseMicrotypeSet[protrusion]{basicmath} 
}{}
\makeatletter
\@ifundefined{KOMAClassName}{
  \IfFileExists{parskip.sty}{%
    \usepackage{parskip}
  }{
    \setlength{\parindent}{0pt}
    \setlength{\parskip}{6pt plus 2pt minus 1pt}}
}{
  \KOMAoptions{parskip=half}}
\makeatother
\usepackage{color}
\usepackage{fancyvrb}

\DefineVerbatimEnvironment{Highlighting}{Verbatim}{commandchars=\\\{\}}
\newenvironment{Shaded}{}{}

\newcommand{\BuiltInTok}[1]{\textcolor[rgb]{0.00,0.50,0.00}{#1}}

\newcommand{\CommentTok}[1]{\textcolor[rgb]{0.38,0.63,0.69}{\textit{#1}}}

\newcommand{\ControlFlowTok}[1]{\textcolor[rgb]{0.00,0.44,0.13}{\textbf{#1}}}

\newcommand{\KeywordTok}[1]{\textcolor[rgb]{0.00,0.44,0.13}{\textbf{#1}}}
\newcommand{\NormalTok}[1]{#1}
\newcommand{\OperatorTok}[1]{\textcolor[rgb]{0.40,0.40,0.40}{#1}}

\newcommand{\StringTok}[1]{\textcolor[rgb]{0.25,0.44,0.63}{#1}}
\newcommand{\VariableTok}[1]{\textcolor[rgb]{0.10,0.09,0.49}{#1}}

\usepackage{longtable,booktabs,array}
\usepackage{caption}
\usepackage{calc} 
\usepackage{etoolbox}
\makeatletter
\patchcmd\longtable{\par}{\if@noskipsec\mbox{}\fi\par}{}{}
\makeatother
\usepackage{graphicx}
\usepackage{tabularx}
\makeatletter
\newsavebox\pandoc@box
\newcommand*\pandocbounded[1]{
  \sbox\pandoc@box{#1}%
  \Gscale@div\@tempa{\textheight}{\dimexpr\ht\pandoc@box+\dp\pandoc@box\relax}%
  \Gscale@div\@tempb{\linewidth}{\wd\pandoc@box}%
  \ifdim\@tempb\p@<\@tempa\p@\let\@tempa\@tempb\fi
  \ifdim\@tempa\p@<\p@\scalebox{\@tempa}{\usebox\pandoc@box}%
  \else\usebox{\pandoc@box}%
  \fi%
}
\def\fps@figure{htbp}
\makeatother
\providecommand{\tightlist}{%
  \setlength{\itemsep}{0pt}\setlength{\parskip}{0pt}}
\usepackage{bookmark}
\IfFileExists{xurl.sty}{\usepackage{xurl}}{} 
\makeatletter
\@ifundefined{xmpquote}{}{}
\makeatother
\hypersetup{
  pdftitle={Before the First Edit? Measuring When Coding-Agent Guidance Actually Arrives},
  pdfauthor={Thupakula Sai Varun},
  hidelinks,
  pdfcreator={LaTeX via pandoc}}

\title{\vspace{-1.2em}\fontsize{19}{22}\selectfont\bfseries Before the First Edit? Measuring When Coding-\allowbreak{}Agent Guidance Actually Arrives\vspace{-0.55em}}
\author{Thupakula Sai Varun\\[0.2em]{\small Independent Researcher, Hyderabad, India}\\{\small Correspondence: \href{mailto:saivarun1410@gmail.com}{saivarun1410@gmail.com}}\\{\small ORCID: \href{https://orcid.org/0009-0006-4314-7055}{0009-0006-4314-7055}}}
\date{}

\begin{document}
\maketitle

\section{Abstract}\label{abstract}

When a coding-agent harness promises guidance before the first source
edit, what event does it actually intercept? We audit five separate
events: policy assignment, request eligibility, trigger, plan delivery,
and observed source-content change. A prescribed-action comparison
exposes the coverage of three gate rules but cannot tell us how an agent
would continue after receiving a plan. We therefore run local agents on
public software issues and hash Python files before and after each
executed action. In a pilot-guided, 76-arm study using Qwen2.5-Coder
14B, 15 arms change Python source across seven tasks and three
repositories. Guidance precedes every observed first change, but every
change uses a structured editor and the study misses its prespecified
20-arm coverage target. Post hoc local tests find seven changed arms
that pass the tested target and regression selectors; four of those
seven use incomplete regression lists. Public PAUSE and SI2CA
trajectories add source-text changes bracketed by adjacent reads, but
lack synchronized filesystem states and cannot identify exact first
mutations. The audit makes the timing claim testable within a stated
source scope. These data do not show that planning improves repair
correctness or covers shell-first writes in live agent runs.

\textbf{Keywords:} coding agents; agent harnesses; empirical software
engineering; intervention fidelity; trace analysis; software agent
evaluation.

\section{1. Introduction}\label{introduction}

A coding agent does not edit a repository directly. It asks a harness to
run a tool. The harness can stop that request, ask for a plan, and
return the plan to the agent before work continues. The timing claim
sounds simple: the plan arrives before the first source edit. In
practice, the harness must decide which requests count. A rule that
watches structured editor calls may never see a Python script, a copy
command, or a shell pipeline that changes the same file. Even when the
rule fires, the Planner may fail before its advice reaches the agent.

We study the gap between the rule a system implements and the source
changes a reader might assume it covers. Figure 1 shows the events we
keep separate. A final patch can show that a file changed, but it
usually cannot place the first change before or after a particular tool
call. For that, we need ordered guidance events and source-state
observations around execution.

\begin{figure*}[t]
\centering
\pandocbounded{\includegraphics[keepaspectratio,alt={The five
observations in a boundary audit: assignment, configured eligibility,
recorded trigger, delivered guidance, and an observed source-content
delta. The dashed path indicates that a source change can occur through
an action outside the configured gate. OpenAI Codex assisted in
rendering the author's diagram; the author checked it against the study
definitions.}]{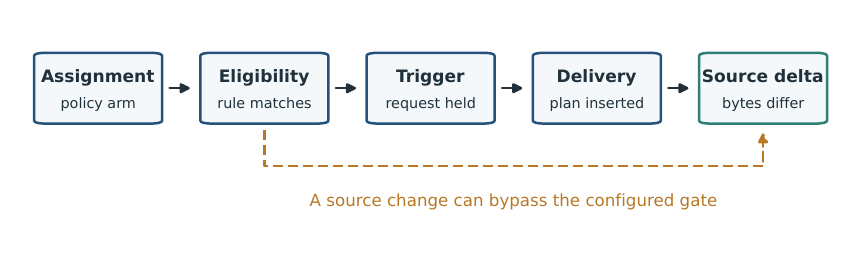}}
\caption{The five observations in a boundary audit: assignment,
configured eligibility, recorded trigger, delivered guidance, and an
observed source-content delta. The dashed path indicates that a source
change can occur through an action outside the configured gate. OpenAI
Codex assisted in rendering the author's diagram; the author checked it
against the study definitions.}\label{fig:boundary-model}
\end{figure*}

\href{https://arxiv.org/html/2610.29870v1}{PAUSE (Patel et al., 2026)}
is an unusually useful setting for studying this distinction. It
releases per-task, step-level traces for multiple harnesses, reports
positive aggregate outcomes, defines an editor-request trigger, and
explicitly notes that shell changes can bypass the Pi trigger. Its
\href{https://huggingface.co/datasets/pause-fse27-artifacts/PAUSE-trajectories}{dataset
card} also specifies that Verified SWE-agent/Pi and OpenHands use
different source-path rules. We use these public data to measure
delivery and trace-visible bypass patterns. We make no claim that a plan
would have improved a run in which it was not delivered; that
counterfactual would require an actual controlled continuation.

\href{https://github.com/Self-Improving-Coding-Agents/SI2CA/blob/main/reports/SI2CA_Technical_Report.pdf}{SI2CA
(SI2CA Team, 2026)} supplies a second, distinct runtime policy: its
discovered strategy branches early agent turns until a third
mutation-like command. The released source implements that boundary with
literal command markers, and the
\href{https://huggingface.co/datasets/Self-Improving-Coding-Agents/SI2CA-Training-Trajectories}{public
trajectories} include per-turn gate decisions. This allows a separate
check of lexical boundary coverage in a bash-only harness. It does not
turn SI2CA into an independent PAUSE replication; the interventions and
estimands differ.

The paper makes three contributions. First, it gives a reproducible way
to separate gate eligibility, trigger recording, plan delivery, and a
confirmed file-content change. Second, it compares gates on constructed
actions and in three separate local multi-turn cohorts, using source
hashes before and after each executed action. The last cohort uses a
larger model and stochastic repeats to test whether the near-identical
earlier runs were a property of the setup. Third, it revisits PAUSE and
SI2CA public traces to show what can, and cannot, be established when
such hashes are missing. Our own experiments also expose limits: source
changes come from few tasks, all live first changes use a structured
editor, and no complete repair outcomes were graded. We keep those
limits beside the positive timing observations.

\section{2. Related work}\label{related-work}

\subsection{2.1 Interventions depend on the tool
interface}\label{interventions-depend-on-the-tool-interface}

\href{https://arxiv.org/html/2610.29870v1}{PAUSE (Patel et al., 2026)}
triggers planning at a source-editor request and notes that Pi can also
edit through the shell. Its released trajectories make it possible to
count recorded triggers and plan artifacts, but they do not include a
filesystem snapshot at each tool boundary.
\href{https://arxiv.org/abs/2609.20804}{Fan et al.~(2026)} show more
broadly that the available action space changes agent behavior.
\href{https://arxiv.org/html/2610.04375v1}{Yang et al.~(2026)} examine
what happens between an agent's intended tool call and the executed
action. This matters when reading Pi's editor errors: an invalid request
in the trace does not, by itself, reveal whether a transport layer
changed the request.
\href{https://catradar-coding-agent.github.io/static/pdfs/CATRadar.pdf}{CATRadar
(Xie et al., 2026)} analyzes behavior patterns in coding-agent training
trajectories. We use traces for a different unit of analysis: the time
and scope of a runtime gate.

\href{https://github.com/Self-Improving-Coding-Agents/SI2CA/blob/main/reports/SI2CA_Technical_Report.pdf}{SI2CA
(SI2CA Team, 2026)} branches early agent turns using a count of
mutation-like commands. Its
\href{https://github.com/Self-Improving-Coding-Agents/SI2CA/blob/e750e3512e4368e4e6a3d3a7392a73dc50bad12a/si2ca/runtime/protocol.py}{released
implementation} uses literal command markers. We inspect that rule on
its own terms; it is a useful second example of how an operational
boundary can differ from a filesystem change.

\subsection{2.2 Process measures and causal
claims}\label{process-measures-and-causal-claims}

\href{https://arxiv.org/abs/2606.16988}{Oderinwale (2026)} studies agent
procedure and variation between runs under the same setup.
\href{https://arxiv.org/html/2610.00372v1}{Xiao et al.~(2026)} use
common states before branching to study individual recovery.
\href{https://arxiv.org/html/2608.22960}{He et al.~(2026)} distinguish
action, task, and step effects. Together, these results explain why a
task-matched PAUSE pair, generated as two separate trajectories, cannot
tell us what a plan would have done at one fixed moment in one run. Our
arm-level outcome summaries remain descriptive; the boundary audit
measures delivery and source state.

\subsection{2.3 Audit trails and write
observation}\label{audit-trails-and-write-observation}

Structured agent logs and pre-execution mediation already exist in
\href{https://arxiv.org/html/2605.13357}{Zhong and Zhu (2026)} and
\href{https://arxiv.org/html/2604.05485}{Nian et al.~(2026)}.
\href{https://github.com/Vlad-Mirzoian/retrace/blob/main/README.md}{Mirzoian
(2026)} identifies potentially untracked shell writes, while the
\href{https://github.com/dreadnode/agent-lens/tree/9eab2b0ca729947cbfed3ecd8cab3f45521447cf}{AgentLens
harness} records step-level changes through shadow Git. These works
establish logging and write tracking as available techniques. Our method
joins those observations to a specific intervention rule and asks
whether a delivered plan actually precedes the first observed source
change.

A rule must also be interpreted using its real configuration.
\href{https://arxiv.org/html/2610.07639v2}{Zhu et al.~(2026)} document a
Python-interpreter route around a command gate, and
\href{https://arxiv.org/html/2609.07360v3}{Kapner et al.~(2026)} show
how a scanner can misread client permission semantics. For PAUSE, an
inactive prefix list in the paper source is consistent with several
trace decisions, but it does not authenticate the historical live rule.
Finally, three inspected
\href{https://huggingface.co/datasets/trace-commons/agent-traces/tree/112ebd4d03ce852b00e935d523107c3d0c9a65bf}{Trace
Commons} sessions include backup metadata without recoverable bytes;
they cannot supply the missing first-mutation observations for PAUSE or
SI2CA.

\section{3. Questions We Can Test}\label{questions-we-can-test}

\textbf{RQ1: Owned gate fidelity.} On prescribed single actions, how
often do editor-only, editor-plus-lexical, and all-request gates deliver
guidance before known package-source changes, and how many actions
without an immediate delta receive guidance? In multi-turn agent runs,
how often does each of four gates deliver a real plan before the first
observed Python-content change, and what guidance timing, token, and
latency counts accompany that delivery?

\textbf{RQ2: PAUSE delivery coverage.} In each PAUSE online setting, how
often is the plan inserted, and how often do unexposed runs nevertheless
end with a modified existing source file? In Pi no-trigger traces, how
often do failed direct editor calls precede a visible shell-write
attempt on the same modified existing file? Do not call a shell command
a successful mutation solely because it returned without error.

\textbf{RQ3: Transfer to another intervention rule.} Does SI2CA's public
early-commit command predicate count explicit candidate writes in
independent agent traces, and what does its per-turn gate log show when
it does not? Keep lexical coverage separate from actual first-write
recall and treatment effects.

\textbf{RQ4: Implications for evaluation.} What claims do task-paired
arm outcomes support, and which trajectory-level claims require shared
state or repeated continuations? Can delivery, path eligibility, and
pre-intervention comparability be reported without conditioning a causal
effect estimate on post-assignment behavior?

\section{4. Data, Agent, and
Measurement}\label{data-agent-and-measurement}

\subsection{4.1 Data and provenance}\label{data-and-provenance}

The prescribed-action material consists of pinned base-commit Python
files from marshmallow, pydicom, and astroid, selected from the public
SWE-bench Lite development split. Each action and policy pair runs in a
fresh temporary workspace. The multi-turn extension uses all 23
development-split issues across six repositories: marshmallow, pvlib,
pydicom, astroid, pyvista, and sqlfluff. Our code specifies eligibility,
guidance insertion, action execution, and state observation; no PAUSE
configuration is used. Secondary data comprise the 500-task SWE-agent
Verified PAUSE/baseline pair, 300-task SWE-agent Multilingual pair,
500-task Pi Verified pair, and 500-task OpenHands Verified pair from the
pinned public trajectory release. SI2CA's standard and
discovered-strategy arms supply an additional lexical coverage check
using its pinned implementation and released per-turn gate logs.
SWE-rebench five-run configurations supply a contextual repeatability
reference, subject to their separate rights and provenance. Source
revisions, hashes, and historical repository license notices are
recorded in the reproduction package.

\subsection{4.2 Definitions}\label{definitions}

\begin{itemize}
\tightlist
\item
  \textbf{Assignment:} a task is run under the PAUSE condition.
\item
  \textbf{Recorded trigger:} the released \texttt{paused=true} field.
\item
  \textbf{Stored plan artifact:} a saved plan, identified by a nonempty
  \texttt{plan\_as\_inserted} field. We report it separately from the
  trigger because two triggered rows lack it, and separately from
  planner-token accounting because three Pi rows have an artifact with
  zero reported tokens. In all 260 triggered Pi rows with an artifact,
  the exact text also appears once as a user message in the released
  trajectory; this verifies trace placement but not how the model used
  it.
\item
  \textbf{Configured eligibility:} an editor request targeting the
  source set defined by that experiment's harness-specific path rule.
  The exact list of package directories must be obtained or
  reconstructed before a case-level eligibility finding.
\item
  \textbf{Broad source-change proxy:} final diff modifies a pre-existing
  Python file (\texttt{diff\ -\/-git} plus \texttt{-\/-\/-\ a/path}),
  excluding newly created scratch files. This does not prove when that
  file changed or whether the path was eligible under the configured
  rule.
\item
  \textbf{Failed editor call:} structured \texttt{edit}/\texttt{write}
  request linked by tool-call ID to an \texttt{isError=true} result on
  the same final-diff path.
\item
  \textbf{Visible shell-write candidate:} shell command with a
  recognized write primitive and a literal final-diff path, optionally
  requiring no reported error. This is a lower-bound pattern detector,
  not a filesystem oracle.
\item
  \textbf{Successful direct editor action:} a tool result that reports
  success for a matching existing path (or SWE-agent observation stating
  that the file was edited).
\end{itemize}

\subsection{4.2.1 Intervention-boundary
model}\label{intervention-boundary-model}

For a run, let \texttt{Z} denote assignment to a policy, \texttt{C\_t}
the proposed command or tool request at turn \texttt{t},
\texttt{P(C\_t)} the policy's configured eligibility predicate,
\texttt{T\_t} the recorded trigger, \texttt{A\_t} a stored intervention
artifact, and \texttt{M\_t} a confirmed source-file delta during
execution. The broad engineering promise ``guidance before the first
source mutation'' requires the artifact to reach the agent before the
first \texttt{M\_t}. An editor-request hook instead implements a
narrower pre-execution promise at the first \texttt{C\_t} for which
\texttt{P(C\_t)=true}. These promises coincide only when the predicate
covers every mutation channel and relevant path, and when triggering
reliably produces usable guidance before execution.

The released PAUSE traces observe \texttt{C\_t}, \texttt{T\_t}, and a
stored \texttt{A\_t}; they have final patches and tool responses but not
a per-turn \texttt{M\_t} state stream. The SI2CA release additionally
exposes its per-turn branch/gate decision, but still lacks synchronized
\texttt{M\_t}. We therefore report \emph{candidate} source writes
inferred from command form, path, return code, and final diff, never
substitute them for confirmed deltas. This observability model yields
distinct failure or scope modes:

\begin{itemize}
\tightlist
\item
  \textbf{Predicate scope:} A source path may be edited without a PAUSE
  trigger. The broad source-change proxy can exceed the configured
  scope; exact eligibility needs the path configuration.
\item
  \textbf{Tool channel:} An invalid direct editor call followed by a
  shell or Python write candidate shows a route outside an editor hook.
  The candidate's exact filesystem effect remains uncertain.
\item
  \textbf{Trigger to artifact:} \texttt{paused=true} with empty
  \texttt{plan\_as\_inserted} shows that a recorded trigger need not
  have a stored plan artifact.
\item
  \textbf{Lexical counter:} An SI2CA write candidate without a listed
  command marker does not increment the published substring counter.
\item
  \textbf{State observation:} Without a per-turn filesystem digest, the
  released trace cannot establish the first actual mutation or whether a
  candidate was a no-op.
\item
  \textbf{Outcome attribution:} Independently rerun task arms cannot
  supply an individual same-state plan effect from one task-paired
  outcome difference.
\end{itemize}

\subsection{4.3 Study design and
uncertainty}\label{study-design-and-uncertainty}

The design combines a controlled gate comparison with a secondary
observational trace audit; it is not a randomized experiment of agent
outcomes. The 39 constructed action cells intentionally contain
difficult write forms and are not sampled from a population of agent
actions. Denominators for released data are their finite cohorts; report
exact counts and proportions. Where generalization beyond a cohort is
discussed, account for clustering by repository and the small number of
repositories. Do not condition an average-effect estimate on delivery or
final patch, both of which are post-assignment. Report unknown
classifications. The author set the research direction, reviewed
successive drafts and results, and requested further validity checks.
OpenAI ChatGPT and Codex supported literature discovery, analysis-code
development, preliminary trace review, figure preparation, and
manuscript drafting. The two selected 20-case trace samples received
tool-assisted preliminary checks, not independent human coding; no
inter-rater reliability is claimed.

We froze the controlled validation protocol before executing its suite.
It selects one base-commit Python file from each of three public
SWE-bench Lite development repositories and runs 13 specified actions in
fresh temporary workspaces per repository. Direct before/after target
bytes establish the declared package-source outcome; a separate
per-action \texttt{.py} hash inventory supplies the observer decision.
The published SI2CA marker tuple is evaluated for shell actions only.
The suite includes editor success, rejection, and no-op; copy, move,
inline Python, helper execution, Perl in-place editing, \texttt{tee}
write and no-op, a partial write with nonzero exit status, read-only
action, and a new test file outside the declared package-source set. The
cells deliberately stress known ambiguities and are not a probability
sample of agent behavior. Its executable, per-cell results, and
limitations are in \texttt{analysis/controlled-boundary-review.md}.

A second, pre-execution protocol applies three locally implemented rules
to the same 39 cells: editor-only, editor plus nine locally fixed
literal command markers drawn from SI2CA's public rule, and all tool
requests. Each policy/action pair starts in a fresh workspace. The
harness records its eligibility decision and inserts a deterministic
placeholder guidance artifact before an eligible action executes, then
compares source bytes and workspace hashes after execution. Every rerun
must match the original controlled cell's command, exit code,
target-byte hashes, and changed paths. The external SI2CA source check
is optional provenance verification; the executable gate uses the marker
tuple declared in our package. The artifact tests delivery timing only;
it does not advise or influence an agent. The all-requests rule is a
coverage upper bound that knowingly triggers on reads and no-ops.

We separately prespecified a multi-turn local-agent study on all 23
SWE-bench Lite development tasks, spanning six repositories. Its v2
amendment was committed after one failed diagnostic arm and before any
v2 output. Each task/policy arm begins at the same pinned base commit
and receives only the issue text, not the reference patch. The locally
installed Qwen2.5-Coder 7B model proposes one JSON tool action per turn
from structured reads/editors and a sandboxed shell. Four policies are
compared: editor-only, editor plus literal command markers, any tool
with write capability (editor or shell), and every non-finish request.
On the first eligible proposed request, the harness holds the action,
obtains a task-specific plan from the same local model, inserts it into
the Executor conversation, and requires the Executor to choose a new
action. Thus the guidance can affect subsequent behavior, unlike the
placeholder in the 39 cells. Every executed action is bracketed by
hashes of all regular Python files in the temporary repository. The
first observed Python-content delta is ordered relative to guidance;
tool outputs, policy decisions, and plan calls are logged. Runs are
capped at eight executed actions and 90 seconds, with all timeouts and
invalid responses retained. The experiment has one run per task/policy,
no official SWE-bench grading, and no power to estimate a general
task-success effect. A separate action replayer rebuilds each base
commit and checks every tool result and source-hash transition. A
source-minimal bundle retains executable actions but omits raw model
conversations and plan text; it supports action replay, not independent
authentication of model generation.

After the development cohort, we froze an \textbf{adaptive follow-up} on
30 disjoint SWE-bench Lite test tasks from five repositories. A
deterministic hash rule selected the tasks before follow-up inference.
Each task has two fixed seeds under each of the four policies: 240
planned arms. Policy order rotates across tasks and seeds. The agent is
instructed to inspect the checkout before editing; this first-action
request is advisory, not a hidden constraint. The same local model,
tools, write-confined shell, eight-action and 90-second limits,
task-specific Planner, and regular-\texttt{.py} before/after hashes
apply. Temperature remains zero. One malformed-action diagnostic led to
a parser-normalization amendment before any v2 follow-up output; all
reported v2 arms share that parser. The predeclared feasibility
condition requires complete replay-valid records under every policy and
at least 20 mutation-bearing arms in three repositories. The protocol
says ``30 task/seed pairs,'' an arithmetic label error: 30 tasks under
two seeds yield 60 pairs. We preserve the frozen rule and report all 240
arms, including no-action and timeout stops. The test split is not a
guarantee against prior model exposure. No official issue grading or
task-success claim is made.

After the 240-arm cohort missed its mutation-coverage target, we froze a
separate small-patch cohort. A model-selection amendment was recorded
before its main runs. Its 19 test-split tasks span four repositories and
have a reference patch with at most 20 added/deleted lines on one
existing Python path. They exclude the earlier follow-up and three pilot
tasks. The disjoint pilot favored a locally served Qwen2.5-Coder 14B
with an action-envelope JSON Schema over the tested 7B settings, but
this adaptive choice may favor the selected task frame. The 14B Executor
samples at temperature 0.7 with task-derived paired seeds; the same
model supplies short plans. Each task receives two repeats under
editor-only and all-request gates, in alternating policy order: 76
planned arms. The tool and source observer are unchanged, while the
limits increase to 16 actions, 24 turns, and 300 seconds. The runner
checks the pinned issue and model-manifest hashes and stops before an
arm if free disk falls below 6 GiB. We specified a task-cluster
descriptive bootstrap within the four fixed repositories before
inspecting the main results. Repeats are grouped by task; the intervals
describe this selected frame, not a population of repositories or repair
tasks. The original 20-changing-arms/three-repositories feasibility
target remains unchanged. The 14B configuration and results are not
pooled with either earlier cohort.

\subsubsection{Agent and policy
configuration}\label{agent-and-policy-configuration}

Within each local cohort, one model plays both roles: Qwen2.5-Coder 7B
in the first two cohorts and Qwen2.5-Coder 14B in the small-patch
cohort. As Executor, it returns one JSON action at a time and can list,
search, read, replace, or write repository files; it also has a
restricted shell. As Planner, it sees the issue and recent tool history
and returns a short text plan. The Planner is called only once, at the
first eligible request. The held request is discarded; the Executor sees
the plan and chooses a new action. The reference patch is withheld from
both roles. A policy comparison changes \emph{when} the same model
within that cohort receives guidance. Table 1 collects the fixed 7B
settings; the 14B changes are specified above. Appendix A gives the gate
predicate and logging sequence.

\begin{table*}[t]
\centering
\footnotesize
\setlength{\tabcolsep}{4pt}
\renewcommand{\arraystretch}{1.14}
\caption{Fixed local-agent configuration for the two 7B multi-turn
cohorts. Token caps are model output limits per call, not measured
usage.}\label{tbl:model-settings}
\begin{tabularx}{\textwidth}{@{}>{\hsize=0.6500\hsize\raggedright\arraybackslash}X>{\hsize=1.3500\hsize\raggedright\arraybackslash}X@{}}

\toprule
Component & Setting \\
\midrule

Model and server & Qwen2.5-Coder 7B through local Ollama chat API \\
Sampling & Temperature 0; task-derived seed; Planner uses seed + 1 \\
Executor output & One JSON action; cap 700 output tokens per reply \\
Planner output & 3--5 sentence plan; cap 220 output tokens \\
Tools & Five structured file tools, restricted shell, \texttt{finish} \\
Shell boundary & Allowlisted commands; no network, installs, absolute
paths, or compound commands \\
Run limits & At most 8 executed actions, 12 turns, and 90 seconds \\
Source observer & SHA-256 map of regular \texttt{.py} files before and
after each action \\
\bottomrule
\end{tabularx}
\end{table*}

The four strategies differ only in eligibility. \textbf{Editor} holds
\texttt{replace\_text} or \texttt{write\_file}. \textbf{Literal} also
holds a shell request containing one of nine fixed mutation markers.
\textbf{Capability} holds any shell request because the shell can write,
even if this particular command reads. \textbf{All} holds every
non-\texttt{finish} request. A Planner timeout produces no delivered
guidance and stops that arm. Figure 2 shows the action order used to
decide whether a plan precedes the first observed source delta.

\begin{figure*}[t]
\centering
\pandocbounded{\includegraphics[keepaspectratio,alt={Local multi-turn
procedure. The first eligible request is held; a delivered plan returns
the Executor to action selection. Other requests execute, with source
hashes checked before and after each action. A separate replayer checks
the recorded transitions. OpenAI Codex assisted in rendering the
author's workflow; the author checked it against the
protocol.}]{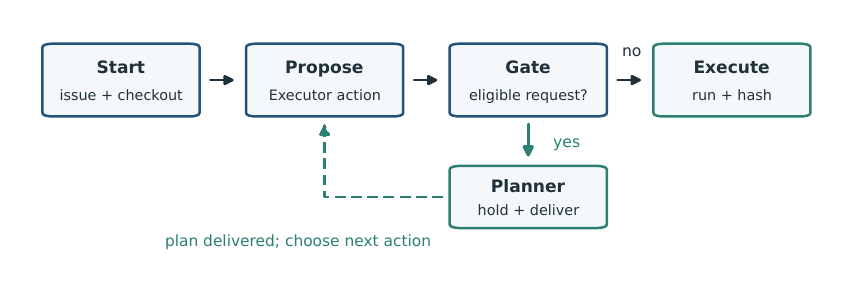}}
\caption{Local multi-turn procedure. The first eligible request is held;
a delivered plan returns the Executor to action selection. Other
requests execute, with source hashes checked before and after each
action. A separate replayer checks the recorded transitions. OpenAI
Codex assisted in rendering the author's workflow; the author checked it
against the protocol.}\label{fig:agent-workflow}
\end{figure*}

\subsubsection{Checks before and after
execution}\label{checks-before-and-after-execution}

Before each arm, the runner verifies the pinned dataset hash and task
manifest, extracts the recorded base commit into a fresh workspace, and
records an initial Python-source hash map. Before an eligible action is
held, it snapshots the workspace; after the Planner call it checks that
planning itself did not change source. Before and after every executed
action, it hashes each regular \texttt{.py} file and records the changed
paths, tool result, and exit code. The first nonempty difference is the
observed first durable Python-source delta. After an arm, independent
replay rebuilds its base commit and checks action order, results, source
hashes, and final state. The follow-up has 236/240 exact raw-output
replays; four network-denied \texttt{pip} retry outputs differ in text,
while a separate scoped check matches their action order, exit codes,
Python-source hashes, and final state. The distinction is retained in
all reproducibility claims. Appendix B lists the actual verification
gates and the limits of this observer.

\subsection{4.4 Trace reconstruction}\label{trace-reconstruction}

For PAUSE, join the treatment and baseline releases on the unique task
ID only for arm-level bookkeeping and pre-trigger comparability
diagnostics. Read \texttt{paused} and planner fields directly, then
inspect the treated run's ordered trace. In Pi, pair assistant tool
requests with results by tool-call ID. An editor request is counted as
successful only when its paired result is not an error and explicitly
reports a write; an \texttt{isError} result never counts as a successful
direct edit. Parse each final Git diff block and require
\texttt{-\/-\/-\ a/\textless{}path\textgreater{}} for a pre-existing
Python file; a newly added reproduction script is not source-change
evidence under this proxy. Literal path matching has token boundaries
and accepts repository-relative, \texttt{./}, and \texttt{/testbed/}
forms. The original shell detector requires a recognized write
primitive; copy/move destinations and inline or helper-script Python
writers are separately declared sensitivity analyses. A clean command
result is a candidate witness, not a file-delta observation. The
mechanism scripts deliberately avoid reading \texttt{resolved}.

For a stricter Pi source-state check, require a successful read-only
bash command immediately before and after a clean same-path shell-write
candidate in a \texttt{paused=false} run. The two read commands must be
identical and their tool-call and result messages strictly ordered;
their outputs must differ without reported errors or truncation. Also
check whether an earlier same-path structured editor request returned an
error. This filters for an observed source-text change interval,
although the exact configured PAUSE path eligibility and first
filesystem mutation remain unavailable.

For triggered Pi runs, locate the inserted plan by requiring exactly one
user message whose content equals \texttt{plan\_as\_inserted}. Count
linked shell-write candidates and successful direct editor actions
before that message only when their target is an existing Python path in
the final diff. The strict source-change check compares identical reads
where available. In a follow-up audit designed after inspecting one
SymPy case, we align successful, bounded structured \texttt{read}
outputs by requested line offset for final-diff existing Python paths
before the plan. We remove continuation footers, require an explicit
limit of 5--150 lines and at least five overlapping returned lines, and
compare the shared lines. This post hoc overlapping-read audit has
unknown recall and is reported separately from the strict identical-read
census.

For a path-scope sensitivity check, fetch the SHA-256-pinned
\href{https://export.arxiv.org/src/2610.29870v1}{PAUSE arXiv TeX
source}, extract the Verified prefix list from an inactive
\texttt{\textbackslash{}commentout} implementation-details block, and
test it against normalized released \texttt{paused\_path} fields. Pi has
no populated \texttt{paused\_path}, so for triggered rows with a unique
plan message we use the immediately preceding
\texttt{edit}/\texttt{write} request only when its tool-call ID links to
a result explicitly saying the batch was not executed because of PAUSE.
Also compare the candidate list with successful direct-editor actions on
modified existing Python files in no-trigger SWE-agent and Pi runs. The
list's inactive status prevents us from identifying it as the live
configuration; the test assesses consistency with observed decisions.

The SWE-agent/OpenHands task-aligned comparison is independently
re-parsed with a stricter verifier that requires the editor action
target, success-observation path, and old-side Python final-diff path to
agree exactly. It reproduces the full 500-task trigger contingency and
every one of the 22 same-path witnesses. This checks extraction
consistency, not the historical path predicate or the filesystem effect
of an editor response.

For SI2CA, we pinned both the dataset revision and the source commit
that defines its literal mutation markers and early-commit window. We
verified the source SHA-256, parsed the marker tuple, and applied the
same substring rule to each executed shell command. A candidate miss had
to name an existing Python file that also appeared in the final diff. We
recognized copy and move destinations, Perl in-place edits, and Python
one-liners that opened a file for writing. We excluded truncated patch
records, which could hide target paths. In the discovered-strategy arm,
we linked each winning command to its executed tool call, return code,
and logged gate phase. Two empty-command representations were normalized
as equivalent. The standard-arm result is an external lexical check;
only the discovered arm also exposes the publisher's gate decision.
Exact identifiers, revisions, and parsing checks are in the replication
guide.

For a high-precision trace-state sensitivity check, require the exact
same read-only bash command immediately before and after an
active-window marker-negative candidate, both reads returning zero
without truncation or warning and naming that candidate's modified
existing Python path. Compare returned text, not merely exit codes. A
changed result brackets a source-content difference if no unlogged
writer intervened. This check has deliberately low and unknown recall:
agents do not usually repeat precisely the same read around every write,
and it cannot identify the first mutation or the content hash at the
execution boundary.

The 20-case PAUSE witness sample and 20-case SI2CA active-window sample
use fixed SHA-256 ranking seeds recorded with the scripts. The
preliminary review checked command text, linked results, target paths,
and relevant final diffs or next-gate decisions. These checks can find
parser mistakes and support lexical precision on the sampled cases. They
do not measure actual first-write recall, inter-rater agreement, or
command-to-patch causation. The local digest fixture validates that
particular command forms can change a file while a substring rule stays
false, without implying that every analogous released command did so.

For prospective measurement feasibility, we froze a separate set of
twelve
\href{https://huggingface.co/datasets/princeton-nlp/SWE-bench_Lite}{SWE-bench
Lite development issues} (Jimenez et al., 2024) across three
repositories before inference. Each public issue was presented to a
local \texttt{qwen2.5-coder:7b} agent on an archive of its recorded base
commit, without showing the reference patch. A constrained text-JSON
file-tool interface allowed Python-file listing, literal search, bounded
reading, exact replacement, and overwrite; it did not allow shell
execution or new files. We hashed every regular \texttt{.py} file before
and after each action and independently replayed logged actions from the
base commits. Selection, model, limits, and outcomes were frozen in the
prospective protocol. This diagnostic is not included in PAUSE/SI2CA
counts and does not evaluate issue resolution or an intervention effect.

\subsection{4.5 Estimands and limits of task
pairing}\label{estimands-and-limits-of-task-pairing}

Let \texttt{Y\_i(z,\ u)} denote task \texttt{i}'s outcome under arm
\texttt{z} and a stochastic execution draw \texttt{u}, and let
\texttt{D\_i(1,\ u)} indicate whether the intervention was actually
delivered in a PAUSE-assigned run. The released online comparison
observes \texttt{Y\_i(1,\ u\_i\^{}1)} and \texttt{Y\_i(0,\ u\_i\^{}0)}
for independently generated draws. Averaging their difference across
tasks can estimate an \textbf{arm-assignment contrast} under the study's
comparability assumptions. A particular task's difference is not the
effect of inserting a plan into one fixed trajectory, because
\texttt{u\_i\^{}1} and \texttt{u\_i\^{}0} need not yield the same
pre-edit state. Conditioning on \texttt{D\_i(1,\ u\_i\^{}1)} or on the
final patch selects on events after assignment; the resulting subset
contrast has no simple randomized interpretation. Likewise, a no-trigger
treated run and its baseline can disagree in outcome without showing a
plan effect. Same-state exposure effects require checkpointed branching
or suitably repeated and state-aware continuations, which this secondary
analysis does not contain.

\section{5. What the Audit Found}\label{what-the-audit-found}

\subsection{5.1 Delivery funnel}\label{delivery-funnel}

The \texttt{paused} trigger flag is true in 465/500 SWE-agent Verified,
289/300 Multilingual, 261/500 Pi Verified, and 486/500 OpenHands
Verified runs. A nonempty \texttt{plan\_as\_inserted} artifact is
present in 465, 288, 260, and 486 respectively. Thus two triggered rows
lack a stored plan artifact. In Pi, \textbf{239 runs have no recorded
trigger but 240 have no stored plan artifact}; the corresponding
Multilingual counts are 11 and 12. Every mechanism count below uses the
\texttt{paused=false} cohort, never the larger artifact-absent cohort.
Three other Pi rows have an artifact with zero reported planner tokens,
so token fields alone cannot determine delivery. These measures should
be interpreted with each condition's distinct tool and path predicate.
The Pi contrast is large but does not identify a harness effect because
execution behavior and implementation details differ.

The two trigger/artifact exceptions have different terminal trace
signals. Multilingual task \texttt{preactjs\_\_preact-2896} ends with
\texttt{Exit\ due\ to\ unknown\ error:\ \textquotesingle{}choices\textquotesingle{}}
and reports zero planner tokens. Pi task \texttt{django\_\_django-16527}
reports 3,747 planner input and 339 output tokens but ends with
\texttt{PAUSE\ invalid\ checkpoint\ or\ teacher\ failure;\ stop.}
Neither trace contains an inserted-plan artifact. The Pi message does
not distinguish its two named causes, and the token count is not
evidence that a usable plan reached the Executor. These cases support a
trigger-to-artifact bookkeeping distinction, not a measured
exposure-failure rate beyond the stored field.

Across the 500 shared Verified task IDs, SWE-agent and OpenHands both
record a trigger in 452, only SWE-agent in 13, only OpenHands in 34, and
neither in one. Of the 34 OpenHands-only trigger tasks, 22 SWE-agent
no-trigger runs nevertheless have a successful direct-editor result on
an existing Python file in their final diff. In all 22, that witnessed
path exactly matches the editor path held by OpenHands on the same task.
These 22 are concentrated in three repository groups: 19
\texttt{pytest-dev}, two Matplotlib, and one Flask. The other 12
OpenHands-only tasks, all Django, lack this particular SWE-agent editor
witness. Within the 19 \texttt{pytest-dev} tasks, Pi likewise records no
trigger; 11 runs have a successful direct-editor witness, ten on the
corresponding OpenHands held \texttt{src/\_pytest/} path. These are
independently generated runs, so matching task IDs and target paths
cannot isolate the effect of different trigger predicates or establish
which action first mutated a file. They do make the scope distinction
concrete at the task and file-path level. The released dataset card
describes package-directory rules for SWE-agent/Pi Verified and a
tracked-file rule for OpenHands Verified; the exact live
package-directory list remains unauthenticated.

As a data-integrity check, the PAUSE summary table and delivery figure
reproduce the four released baseline-to-PAUSE solved-task counts: 251 to
325 (Verified SWE-agent), 85 to 129 (Multilingual SWE-agent), 185 to 215
(Pi Verified), and 227 to 311 (OpenHands Verified). These positive
arm-level comparisons coexist with incomplete intervention delivery.
They cannot be read as per-trajectory treatment effects because the
task-matched arms are independent runs.

\begin{table*}[t]
\centering
\footnotesize
\setlength{\tabcolsep}{4pt}
\renewcommand{\arraystretch}{1.14}
\caption{PAUSE trigger, plan artifact, and outcome counts by harness. B
and P denote baseline and PAUSE-assigned
runs.}\label{tbl:pause-delivery}
\begin{tabularx}{\textwidth}{@{}>{\hsize=2.2000\hsize\raggedright\arraybackslash}X>{\hsize=0.8000\hsize\raggedleft\arraybackslash}X>{\hsize=0.8000\hsize\raggedleft\arraybackslash}X>{\hsize=0.8000\hsize\raggedleft\arraybackslash}X>{\hsize=0.8000\hsize\raggedleft\arraybackslash}X>{\hsize=0.8000\hsize\raggedleft\arraybackslash}X>{\hsize=0.8000\hsize\raggedleft\arraybackslash}X@{}}

\toprule
Cohort & Runs & Trigger & Plan & No trigger & Existing \texttt{.py} diff* & Solved B → P \\
\midrule

SWE V & 500 & 465 & 465 & 35 & 23 & 251 → 325 \\
SWE M & 300 & 289 & 288 & 11 & 0 & 85 → 129 \\
Pi V & 500 & 261 & 260 & 239 & 136 & 185 → 215 \\
OH V & 500 & 486 & 486 & 14 & 1 & 227 → 311 \\
\bottomrule
\end{tabularx}
\end{table*}

\emph{SWE denotes SWE-agent; OH denotes OpenHands. V and M denote
Verified and Multilingual. The existing \texttt{.py} diff column is
restricted to runs with no recorded trigger.}

\begin{figure*}[t]
\centering
\pandocbounded{\includegraphics[keepaspectratio,alt={PAUSE trigger and
stored-plan funnel across four cohorts. The marked no-trigger
source-diff segment uses a broad final-diff proxy; small
trigger-without-artifact segments are included. Counts are in the
adjacent PAUSE summary table. OpenAI Codex assisted with the rendering
code and layout; the author checked the figure against that
table.}]{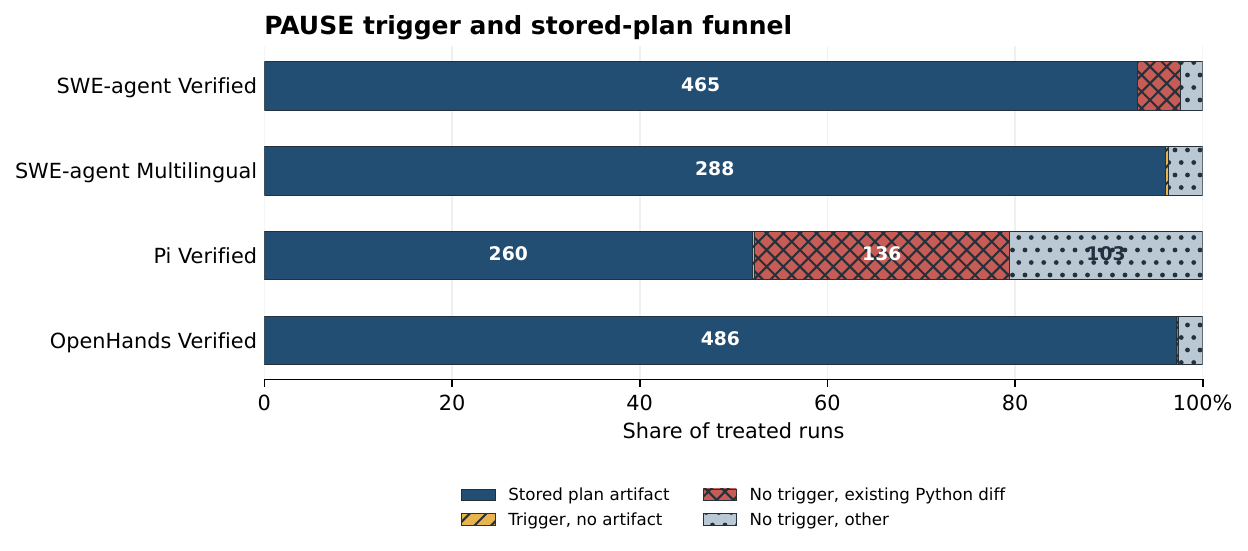}}
\caption{PAUSE trigger and stored-plan funnel across four cohorts. The
marked no-trigger source-diff segment uses a broad final-diff proxy;
small trigger-without-artifact segments are included. Counts are in the
adjacent PAUSE summary table. OpenAI Codex assisted with the rendering
code and layout; the author checked the figure against that
table.}\label{fig:pause-funnel}
\end{figure*}

In Pi, all 260 nonempty plan artifacts in triggered rows also appear
exactly once as user messages in the released trace. Among those 260
runs, four have six clean, source-directed shell-write candidates on a
final-diff existing Python path before the plan message; four other runs
have a successful direct editor action on such a path before the plan,
all in \texttt{tests/}. These are selected path and tool observations,
not a prevalence estimate of actual pre-plan mutation. The strict
identical-read test finds no changed pre-plan pair in the four
shell-candidate runs. The post hoc overlapping-read audit finds 94
comparable pre-plan pairs in 35 of the 249 triggered Pi runs with an
existing-Python final diff; four pairs differ, all in one run,
\texttt{sympy\_\_sympy-19040}. Its overlapping reads of
\texttt{sympy/polys/factortools.py} show different text in the anchored
\texttt{dmp\_ext\_factor} function region after three source-directed
shell actions and before plan insertion. The read windows are not
identical, and the trace later contains a checkout and another writer;
this supports a pre-plan source-change interval but not atomic
attribution to one action. The case limits the broad ``before any source
change'' interpretation without contradicting PAUSE's narrower
editor-request trigger contract.

\textbf{Final-diff extension sensitivity.} The table's existing
\texttt{.py} column is a broad source proxy, not the experiment's
configured eligibility predicate. An all-extension inventory finds only
four additional no-trigger runs with a modified existing file, all in Pi
and all changing documentation or metadata (\texttt{README.md},
\texttt{AUTHORS}, or \texttt{doc/*.rst}). No other existing
source-language extension appears in the no-trigger final diffs. This
does not detect intermediate changes later reverted or establish
configured path eligibility. The OpenHands case edits a test path
(\texttt{django/test/testcases.py}), which its dataset card says the
tracked-file trigger excludes.

\subsection{5.2 Editor failure and shell fallback in
Pi}\label{editor-failure-and-shell-fallback-in-pi}

Among 239 Pi runs with no recorded trigger, 213 have a final diff and
136 modify an existing Python file. In 128 of those 136, a direct editor
request on a modified existing path returns an error; 125 show
\texttt{edits.0:\ must\ be\ object}, consistent with a string supplied
where the tool expects an object. In 90, an errored editor call precedes
a same-path shell-write candidate; 76 have at least one later same-path
candidate without a reported error. The 20-case first-pass inspection
found that six first matching shell commands actually reported errors,
which motivates the stricter indicator. The observed sequence is
compatible with editor failure and fallback to shell, but exact first
filesystem mutations are not established from final diffs and command
text alone.

The adjacent-read check supplies stronger evidence for a narrow subset:
nine clean shell-write commands across seven no-trigger Pi runs have
identical successful reads of the same existing Python path immediately
before and after, returning different source-code text. All seven runs
have an earlier same-path errored editor request. These witnesses span
only \texttt{django} (three runs) and \texttt{sympy} (four runs), and
two SymPy runs contribute two qualifying commands each. The preliminary
inspection recorded source-code differences in all nine read pairs; a
separate parser independently confirmed the nine selected
tool-call/result sequences and changed read text. The audit encounters
211 clean recognized same-path write candidate commands, but only these
nine have qualifying identical adjacent reads. The other 202 lack the
observation pattern; they are not observed no-ops. This is a lower bound
under a strict observation pattern, not a prevalence estimate for all
239 no-trigger runs or proof that the path met PAUSE's hidden
package-directory rule. The command is the sole logged agent action
between the two reads; unlogged concurrent writes remain a theoretical
alternative.

\begin{figure*}[t]
\centering
\pandocbounded{\includegraphics[keepaspectratio,alt={Ordered
observations in django\_\_django-12713, one of the seven strict Pi
witnesses. The shell command lies between identical reads that return
different source text. The diagram encodes order, not elapsed time or
the exact filesystem mutation instant. OpenAI Codex assisted in
rendering the author's timeline; the author checked it against the trace
record.}]{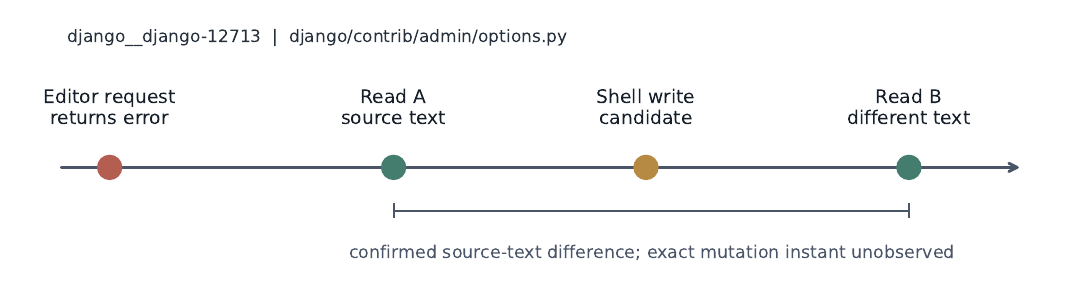}}
\caption{Ordered observations in
\protect\texttt{django\_\_django-12713}, one of the seven strict Pi
witnesses. The shell command lies between identical reads that return
different source text. The diagram encodes order, not elapsed time or
the exact filesystem mutation instant. OpenAI Codex assisted in
rendering the author's timeline; the author checked it against the trace
record.}\label{fig:boundary-timeline}
\end{figure*}

\begin{table*}[t]
\centering
\footnotesize
\setlength{\tabcolsep}{4pt}
\renewcommand{\arraystretch}{1.14}
\caption{Nested and related criteria for PAUSE Pi runs without a
recorded trigger. Rows with different predicates are not all mutually
nested.}\label{tbl:pi-fallback}
\begin{tabularx}{\textwidth}{@{}>{\hsize=1.6000\hsize\raggedright\arraybackslash}X>{\hsize=0.4000\hsize\raggedleft\arraybackslash}X@{}}

\toprule
Nested or related Pi criterion & Runs \\
\midrule

No recorded trigger (\texttt{paused=false}) & 239 \\
Any final diff & 213 \\
Final diff modifies existing \texttt{.py} file & 136 \\
Same-path editor request returns an error & 128 \\
Specific \texttt{edits.0:\ must\ be\ object} error & 125 \\
Errored editor followed by recognized same-path shell write candidate &
90 \\
Same sequence with a later candidate reporting no error & 76 \\
Same-path shell candidate bracketed by changed identical reads & 7 \\
\bottomrule
\end{tabularx}
\end{table*}

The schema-error row is a subset of errored-editor cases but not
necessarily nested in the 90-row sequence. Rows below the final-diff
proxy are post hoc mechanism descriptions, not treatment-effect
subgroups.

The 90 ordered fallback witnesses in the Pi fallback table occur in nine
dataset repository groups: \texttt{django} (40), \texttt{sympy} (21),
\texttt{scikit-learn} (8), \texttt{matplotlib} (6), \texttt{sphinx-doc}
(5), \texttt{pydata} (5), \texttt{pytest-dev} (3), \texttt{pallets} (1),
and \texttt{mwaskom} (1). The mechanism is therefore less concentrated
than the SWE-agent direct-editor pattern, although \texttt{django} and
\texttt{sympy} contribute 61/90 cases. These are repository clusters,
not 90 independent estimates of a general failure rate.

For example, in \texttt{django\_\_django-12308}, three \texttt{edit}
calls on \texttt{django/contrib/admin/utils.py} return the same schema
error before two \texttt{sed\ -i} commands name that file and report no
error; the run has a final diff for that file and no recorded trigger.
This is a concrete ordered fallback witness. It cannot by itself show
whether the first \texttt{sed} command made the final patch, whether the
path was eligible under the hidden package-directory configuration, or
whether a plan would have changed the outcome.

An exploratory visibility taxonomy finds 75 of the 136 no-trigger,
existing-Python-diff runs with a recognizable shell-write candidate and
no successful direct editor action, 14 with a successful direct editor
action (one also later uses shell), and 47 unclassified. A separately
declared sensitivity detector for \texttt{cp}/\texttt{mv} destinations
identifies 22 of those 47 unknowns, leaving 25 unclassified. These are
progressively broader candidate definitions; they must not be described
as proven mutation channels.

A second sensitivity pass identifies 24 of the 25 remaining cases
through explicit Python writers: nine run an inline \texttt{python\ -c}
command with a literal final-diff path, and 19 write a helper script
containing such a path before executing it; four meet both criteria. The
sole unmatched run constructs its source path dynamically inside a
written and executed Python helper. These findings reveal how strongly a
literal shell-command detector undercounts indirect write candidates.
The helper scripts' contents and execution results provide richer
witnesses, but there are no synchronized filesystem snapshots to
establish each command's effect or the exact first mutation.

\subsection{5.3 Package-directory rule and repository
layout}\label{package-directory-rule-and-repository-layout}

In all 19 \texttt{pytest-dev} SWE-agent Verified runs, PAUSE does not
trigger despite a successful edit to an existing
\texttt{src/\_pytest/*.py} file; 11 Pi runs share this pattern, whereas
all 19 OpenHands runs trigger. The release describes package-directory
rules for SWE-agent/Pi and a tracked-file rule for OpenHands. An
inactive arXiv TeX block lists 16 Verified prefixes, including
\texttt{pytest/} but excluding \texttt{src/\_pytest/},
\texttt{lib/mpl\_toolkits/}, and \texttt{src/flask/}. The list fits all
465 SWE-agent \texttt{paused\_path} values and all 260 linked, rejected
Pi editor targets; none of the 22 SWE-agent or 14 Pi no-trigger
successful-editor paths fall inside. Because this block is not rendered
and may be stale, it supports a scope explanation without authenticating
the live rule. Harness differences can also change chosen paths and
action timing.

The candidate list also splits Pi's 90 editor-error-to-shell fallback
runs: 85 name a modified existing Python path inside it and five
outside. Of the 76 with a later candidate reporting no error, 71 are
inside and five outside; all seven strict changed-read runs are inside.
If this list matches the live rule, inside cases point to alternate-tool
coverage on paths the editor predicate accepts, while outside cases also
involve path scope. The categories remain conditional, and most
candidate commands lack confirmed file deltas.

The path-scope table crosses candidate-prefix scope of final-diff
existing Python paths with primary visible action channels in all 136 Pi
no-trigger runs. All 14 successful direct-editor witnesses are
outside-only; 71 of 114 inside-only runs have a recognized shell
candidate, and 43 have neither primary witness. Broader copy, move, and
Python-helper checks find additional routes among the latter. Three runs
modify paths on both sides of the candidate boundary. The matrix does
not identify configured eligibility or first mutations.

\begin{table*}[t]
\centering
\footnotesize
\setlength{\tabcolsep}{4pt}
\renewcommand{\arraystretch}{1.14}
\caption{Candidate prefix scope of final-diff existing Python paths
versus primary visible action channel in 136 no-trigger Pi runs. The
prefix list is inactive author-source evidence, not authenticated live
configuration.}\label{tbl:pi-scope-channel}
\begin{tabularx}{\textwidth}{@{}>{\hsize=1.6000\hsize\raggedright\arraybackslash}X>{\hsize=0.8500\hsize\raggedleft\arraybackslash}X>{\hsize=0.8500\hsize\raggedleft\arraybackslash}X>{\hsize=0.8500\hsize\raggedleft\arraybackslash}X>{\hsize=0.8500\hsize\raggedleft\arraybackslash}X@{}}

\toprule
Final-diff path scope & Shell candidate only & Successful editor present & Neither primary witness & Runs \\
\midrule

Inside only & 71 & 0 & 43 & 114 \\
Outside only & 3 & 14 & 2 & 19 \\
Both & 1 & 0 & 2 & 3 \\
Total & 75 & 14 & 47 & 136 \\
\bottomrule
\end{tabularx}
\end{table*}

The SWE-agent direct-editor pattern is concentrated: 19 of its 22
no-trigger witnesses come from the single \texttt{pytest-dev}
repository; the other three come from Matplotlib (two) and Flask (one).
Thus 22 task-level witnesses are only three repository-level examples of
this pattern. The \texttt{pytest-dev} result is an important coverage
case, but treating its 19 tasks as independent replications of a general
harness failure would overstate the evidence.

Beyond \texttt{pytest-dev}, the other SWE-agent
no-trigger/successful-editor witnesses target two
\texttt{lib/mpl\_toolkits} files in Matplotlib and one
\texttt{src/flask} file. Pi has corresponding Matplotlib and Flask cases
plus one Django \emph{test} file. The latter is a reminder that our
pre-existing-Python proxy is intentionally broader than production
source. The candidate prefix list excludes all these paths, so the
released cases do not support a broken editor hook claim.

\subsection{5.4 Interpretation of task-paired
outcomes}\label{interpretation-of-task-paired-outcomes}

In no-trigger PAUSE subsets, task-paired baseline outcomes can still
disagree with treated-run outcomes (see the outcome-discordance table).
This is expected when conditions are independently rerun. The no-trigger
subset is selected by the treatment execution path, so it is not a
general no-treatment estimate. Repeated-run SWE-rebench data likewise
show outcome variation under a fixed configuration; that fact is
established in prior work. Our contribution should be a concrete audit
of what an \emph{individual} trajectory claim would require: either a
common checkpoint before intervention or appropriately repeated,
state-aware continuations.

\begin{table*}[t]
\centering
\footnotesize
\setlength{\tabcolsep}{4pt}
\renewcommand{\arraystretch}{1.14}
\caption{Task-paired outcome discordance in PAUSE no-trigger subsets.
These are selected descriptive subsets, not causal null
controls.}\label{tbl:discordance}
\begin{tabularx}{\textwidth}{@{}>{\hsize=1.6000\hsize\raggedright\arraybackslash}X>{\hsize=0.8000\hsize\raggedleft\arraybackslash}X>{\hsize=0.8000\hsize\raggedleft\arraybackslash}X>{\hsize=0.8000\hsize\raggedleft\arraybackslash}X@{}}

\toprule
Cohort & No trigger & Discordant pairs & Baseline / PAUSE only \\
\midrule

SWE-agent Verified & 35 & 5 & 3 / 2 \\
SWE-agent Multilingual & 11 & 2 & 2 / 0 \\
Pi Verified & 239 & 16 & 9 / 7 \\
OpenHands Verified & 14 & 1 & 1 / 0 \\
\bottomrule
\end{tabularx}
\end{table*}

These disagreements are a diagnostic that task matching does not create
an individual same-state counterfactual. Because \texttt{paused=false}
is selected by the treated run, the table is not a null-effect test and
does not estimate how often an unmodified harness would disagree with
itself. Separately, in five-run SWE-rebench configurations, 144/1,110,
158/1,110, and 158/1,110 within-task run pairs disagree on the external
grade for Claude Code, a standalone model, and Codex respectively. That
independent corpus provides context for repeated-run variability, not a
new treatment comparison or a pooled rate across unlike configurations.

\subsection{5.5 External lexical coverage
check}\label{external-lexical-coverage-check}

SI2CA's released source code uses literal command substrings to count
``mutation-like'' actions for its discovered early-commit window.
Applying those exact literals to its independent \emph{standard-arm}
trajectories, 473 of 10,240 runs with a complete diff of an existing
Python file contain a marker-negative \texttt{cp}/\texttt{mv},
\texttt{python3\ -c} writer, or \texttt{perl\ -i} candidate naming a
modified path and recording a zero exit code. the SI2CA candidate table
summarizes the two SI2CA arms. This tests the portability of the
measurement concern beyond PAUSE. It does not measure a missed SI2CA
intervention: these are standard-arm runs, the commands need not be the
first writes, and a zero exit code does not prove a delta.

\begin{table*}[t]
\centering
\footnotesize
\setlength{\tabcolsep}{4pt}
\renewcommand{\arraystretch}{1.14}
\caption{SI2CA zero-return marker-negative write candidates by arm.
Active branch and changed-read observations apply only to
discovered-strategy runs.}\label{tbl:si2ca-candidates}
\begin{tabularx}{\textwidth}{@{}>{\hsize=2.2000\hsize\raggedright\arraybackslash}X>{\hsize=0.7600\hsize\raggedleft\arraybackslash}X>{\hsize=0.7600\hsize\raggedleft\arraybackslash}X>{\hsize=0.7600\hsize\raggedleft\arraybackslash}X>{\hsize=0.7600\hsize\raggedleft\arraybackslash}X>{\hsize=0.7600\hsize\raggedleft\arraybackslash}X@{}}

\toprule
Arm & Runs & Existing \texttt{.py} diff* & Candidate & Active branch & Changed reads \\
\midrule

Standard & 10,780 & 10,240 & 473 & Not applicable & Not applicable \\
Discovered strategy & 10,780 & 10,218 & 347 & 82 & 10 \\
\bottomrule
\end{tabularx}
\end{table*}

The standard and discovered rows are separate generated trajectories,
and the last column exists only for the discovered arm. The different
candidate counts do not estimate an intervention effect.

\emph{The existing \texttt{.py} diff column excludes truncated patches;
candidate counts require a zero-return, marker-negative command naming
such a path. Active branch requires the candidate and following turn to
be branched.}

The released \emph{discovered-strategy} arm exposes the gate decision on
each turn. In 347 of 10,218 complete-patch runs with an existing Python
diff, a zero-return marker-negative candidate names such a path; in 82,
that command is itself on a branched step and the following turn is also
branched. Those 10,218 runs span 752 repository groups; the 82
active-window cases span 75 groups and both SWE-rebench-V2 (57 tasks)
and SWE-smith (25 tasks). Sixty-nine of the 75 groups contribute one
case and none contributes more than three. Because the 752 groups
contribute unequal numbers of eligible tasks, 75/752 is a breadth
descriptor rather than a repository prevalence estimate. A deterministic
20-run preliminary trace check records an explicit write form and
matching target path in all 20 selected commands, while preserving
uncertainty about filesystem effects. This is evidence that the
published lexical predicate does not count some visible candidate write
commands even during an active intervention window. It is not evidence
that any such command caused the final diff or that the gate exceeded
its intended number of \textbf{actual} writes. Synchronized filesystem
observations would be needed to make that stronger claim.

Within those active-window candidates, 41 candidate commands occur after
exactly two prior commands matched the published marker tuple and are
followed by another branched turn. They appear in 34 runs. If both prior
marked commands and the candidate each changed a file under a boundary
defined over all files, the candidate would be at least the third actual
mutation while the marker counter still allowed the next branch. That
conditional is not established by the released state: a marked command
may be a no-op, the candidate may also make no delta, and some prior
marked commands target temporary files rather than repository source.
Across all 82 runs, candidate commands occur after zero, one, or two
prior markers in 1, 50, and 34 runs respectively; three runs appear in
both the one- and two-marker groups, so these counts must not be summed
as disjoint run categories.

At the command level there are 114 active-window candidates across the
82 runs (1 after zero markers, 72 after one, 41 after two). For all 114,
the next logged decision retains its phase: before the first counted
marker, after the first, or after the second. The replication scripts
use the dataset's exact phase values. This verifies the lexical
counter's observed behavior on these released decisions. It does not
imply that each command changed a file.

The strict adjacent-read check finds ten of those 114 candidate
commands, in ten distinct repositories, with a successful, untruncated
identical read immediately before and after the candidate and different
returned source text. The other 104 commands lack qualifying adjacent
reads and therefore have no classification from this check. All ten
observed candidates are \texttt{python3\ -c} writers; the preliminary
inspection recorded source-code differences in each returned region, and
a separate parser independently confirmed the ten selected
tool-call/result sequences, unchanged branch phase, and changed read
text. Parsing those ten inline Python programs also identifies a literal
write-mode \texttt{open()} on the exact read/diff path in every case.
\texttt{aio-libs\_\_aiohttp-8014} is one such case: the same
numbered-line read of \texttt{aiohttp/web\_fileresponse.py} at turns 17
and 19 changes after the turn-18 writer, while the next turn remains in
\texttt{branch\_after\_first\_mutation}. The ten cases provide a
conservative trace-observed content-change subset of the 82 runs. They
do not make the other 104 candidate commands no-ops, and an unlogged
concurrent writer cannot be excluded from this release alone.

The published marker tuple includes \texttt{python\ -c} but not
\texttt{python3\ -c}. The ten adjacent-read cases therefore connect a
concrete source-text change interval to a command form the operational
counter does not count. They still lack independent before/after file
digests exactly at the intervening command boundary.

An additional \emph{post hoc} case, \texttt{awdeorio\_\_mailmerge-76},
provides a stronger trace-supported source-change interval. A read at
turn 9 returns nonempty lines from \texttt{mailmerge/\_\_main\_\_.py}.
Turns 11 and 12 run marker-positive writes to temporary files. At turn
13, a marker-negative
\texttt{cp\ /tmp/new\_main.py\ mailmerge/\_\_main\_\_.py} command
reports return code zero; turns 14 and 15 read an empty target and
report its size as zero bytes. Both turn 13 and turn 14 are labelled
\texttt{branch\_after\_second\_mutation}. The ordered reads support that
the target changed between the pre-read and post-read, and the copy is
the visible source-directed action in that interval. They do not replace
synchronized per-action digests or prove whether the earlier temporary
writes changed contents as intended. This selected witness is not an
estimate that all 82 candidate runs changed source.

A controlled local fixture confirms that \texttt{cp} and
\texttt{python3\ -c} can change a source file while that lexical
predicate remains false; a \texttt{tee} positive control both changes
the file and matches the predicate. This establishes detector behavior
on concrete command forms but does not replay or attribute any released
agent run.

A second reference fixture snapshots file digests immediately before and
after seven ordered actions in a disposable repository. It observes a
successful read and a zero-return no-op with no source delta;
\texttt{cp} and an executed Python helper change \texttt{pkg/a.py}
without a SI2CA literal marker; \texttt{tee} changes the file with a
marker; a command that exits 7 still changes the file; and a newly
created test file changes workspace state outside the fixture's declared
production-source set. This demonstrates why command form, exit status,
and the declared source set must be logged separately from an observed
delta. It is a constructed protocol check, not an estimate from
published trajectories.

We additionally call the \textbf{publisher's hash-verified
\texttt{early\_commit\_window} function} on four controlled sequences
while hashing \texttt{pkg/a.py} around each executed command. Three
marker-negative \texttt{cp} writes produce three successive content
deltas while the published gate remains
\texttt{branch\_before\_first\_mutation}; three marker-positive
\texttt{tee} writes produce three deltas and advance the gate to
\texttt{skip\_after\_3\_mutations}. A mixed \texttt{cp}, Python-helper,
\texttt{tee} sequence produces three deltas but only one counted marker.
Conversely, a \texttt{tee} command that rewrites identical bytes
advances the marker counter without a source-content delta. These
constructed cases directly validate that the gate measures mutation-like
command forms, not file-content changes. They do not establish how often
any released run exhibited these sequences.

\subsection{5.6 Controlled validation against known source
states}\label{controlled-validation-against-known-source-states}

The prespecified 39-action suite uses pinned source bytes from
marshmallow, pydicom, and astroid. The per-action state observer agrees
with the separate direct target-byte comparison in all 39 cells: 24
package-source changes and 15 nonchanges. Of 30 shell actions, the
published SI2CA lexical marker identifies six of 21 actual
package-source changes and marks three of nine nonchanges. The fifteen
missed changes are five intentionally selected shell forms repeated
across three repositories; the three false positives are same-content
\texttt{tee} rewrites. A shell action returning exit code 7 still
changed the source file. New test-file creation is a workspace change
but not a package-source change under this declared scope. These
matrices establish behavior on constructed cases, not precision or
recall in agent populations.

\begin{table*}[t]
\centering
\footnotesize
\setlength{\tabcolsep}{4pt}
\renewcommand{\arraystretch}{1.14}
\caption{Controlled action decisions against independently compared
package-source bytes. \protect\texttt{TP}, \protect\texttt{FP},
\protect\texttt{FN}, and \protect\texttt{TN} refer to detecting a
content delta in the selected package file, not a filesystem write
event.}\label{tbl:controlled-validation}
\begin{tabularx}{\textwidth}{@{}>{\hsize=1.6000\hsize\raggedright\arraybackslash}X>{\hsize=0.8500\hsize\raggedleft\arraybackslash}X>{\hsize=0.8500\hsize\raggedleft\arraybackslash}X>{\hsize=0.8500\hsize\raggedleft\arraybackslash}X>{\hsize=0.8500\hsize\raggedleft\arraybackslash}X@{}}

\toprule
Decision rule and unit & TP & FP & FN & TN \\
\midrule

SI2CA literal marker, 30 shell actions & 6 & 3 & 15 & 6 \\
Per-action state observer, all 39 actions & 24 & 0 & 0 & 15 \\
\bottomrule
\end{tabularx}
\end{table*}

\subsection{5.7 Fully specified guidance
gates}\label{fully-specified-guidance-gates}

We executed the same 39 prescribed single actions under each of the
three locally implemented policies, for 117 controlled policy/action
executions, \textbf{not 117 independent agent trajectories}. Every
execution reproduces the original command, exit status, before/after
target bytes, and changed-path set. Guidance insertion precedes
execution for every eligible request. Because each cell contains a
single request, each package-source delta is the cell's first such
change. the constructed-gate table shows coverage versus delivery on
actions without an immediate package-source delta. Editor-only guidance
precedes three of the 24 selected source changes and is inserted on six
nonchanges. Adding the literal command rule reaches nine source changes,
while inserting nine artifacts on nonchanges. Triggering on every tool
request reaches all 24 but also inserts 15 artifacts on nonchanges.
These artifact counts are a proxy for delivery frequency; no Planner
token cost or latency was measured in this controlled experiment. A plan
before a read could still affect later edits in a multi-turn agent, so
the nonchanging-action count is \textbf{not} a count of unnecessary
interventions.

\begin{table*}[t]
\centering
\footnotesize
\setlength{\tabcolsep}{4pt}
\renewcommand{\arraystretch}{1.14}
\caption{Our three pre-execution policies on 39 constructed
repository/action cells. \protect\texttt{TP}: source change with prior
guidance; \protect\texttt{FP}: guidance without an immediate source
change; \protect\texttt{FN}: source change without prior guidance;
\protect\texttt{TN}: neither. The placeholder guidance never changes the
action.}\label{tbl:owned-gates}
\begin{tabularx}{\textwidth}{@{}>{\hsize=2.2000\hsize\raggedright\arraybackslash}X>{\hsize=0.7600\hsize\raggedleft\arraybackslash}X>{\hsize=0.7600\hsize\raggedleft\arraybackslash}X>{\hsize=0.7600\hsize\raggedleft\arraybackslash}X>{\hsize=0.7600\hsize\raggedleft\arraybackslash}X>{\hsize=0.7600\hsize\raggedleft\arraybackslash}X@{}}

\toprule
Policy & TP & FP & FN & TN & Artifacts \\
\midrule

Editor requests only & 3 & 6 & 21 & 9 & 9 \\
Editor or literal command marker & 9 & 9 & 15 & 6 & 18 \\
Every tool request & 24 & 15 & 0 & 0 & 39 \\
\bottomrule
\end{tabularx}
\end{table*}

\begin{figure*}[t]
\centering
\pandocbounded{\includegraphics[keepaspectratio,alt={Constructed-cell
trade-off. The left bars count package-source changes preceded by
guidance (out of 24); the right bars count actions without an immediate
package-source change that still received guidance (out of 15). These
are prescribed actions, not independent agent trajectories. OpenAI Codex
assisted with the rendering code and layout; the author checked the
figure against the recorded
table.}]{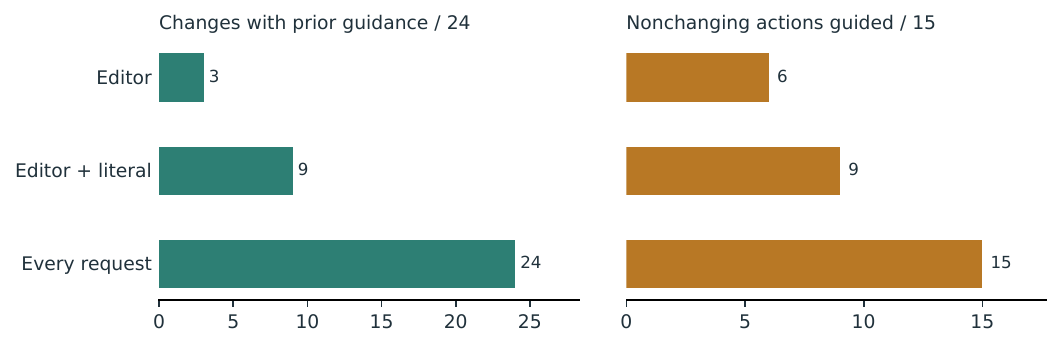}}
\caption{Constructed-cell trade-off. The left bars count package-source
changes preceded by guidance (out of 24); the right bars count actions
without an immediate package-source change that still received guidance
(out of 15). These are prescribed actions, not independent agent
trajectories. OpenAI Codex assisted with the rendering code and layout;
the author checked the figure against the recorded
table.}\label{fig:controlled-tradeoff}
\end{figure*}

\subsection{5.8 Policy eligibility on agent-generated witness
commands}\label{policy-eligibility-on-agent-generated-witness-commands}

As a post hoc cross-corpus check, we applied the four locally specified
predicates to the executed bash requests in the 19 strict changed-read
pairs from PAUSE Pi and SI2CA. These are selected positive witness
intervals in 17 runs, not a held-out probability sample. The writer
command in each interval is bracketed by identical read requests whose
returned source text differs. Eight of nine PAUSE Pi writers match the
literal-marker policy; its remaining \texttt{perl\ -i} writer and all
ten SI2CA \texttt{python3\ -c} writers do not. Editor-only recognizes
none of these shell requests, whereas both the capability and
all-request gates recognize all 19. The capability gate also recognizes
all 38 flanking read \emph{slots} because the publisher's bash tool can
write even when a particular command only reads. Some slots in
multi-pair runs may refer to the same logged read event. These are
counterfactual \textbf{eligibility} decisions on already observed
commands: we did not insert guidance into the publisher runs and cannot
infer their outcomes under our gates. The result shows that
capability-based selectivity depends on the harness exposing distinct
read and write tools; with one general shell, it collapses to
all-request gating for these intervals. The commands, source-text
checks, and decision counts are reproducible by
\texttt{scripts/audit\_witness\_policy\_eligibility.py}.

\begin{table*}[t]
\centering
\footnotesize
\setlength{\tabcolsep}{4pt}
\renewcommand{\arraystretch}{1.14}
\caption{Offline eligibility of four owned policies on selected
publisher-trace events. Read counts are flanking slots with
multiplicity, not distinct requests or measured intervention
burden.}\label{tbl:witness-policy}
\begin{tabularx}{\textwidth}{@{}>{\hsize=1.6000\hsize\raggedright\arraybackslash}X>{\hsize=0.7000\hsize\raggedleft\arraybackslash}X>{\hsize=0.7000\hsize\raggedleft\arraybackslash}X@{}}

\toprule
Policy & Eligible writer commands / 19 & Eligible read slots / 38 \\
\midrule

Editor requests only & 0 & 0 \\
Editor or literal command marker & 8 & 0 \\
Editor or any shell request & 19 & 38 \\
Every tool request & 19 & 38 \\
\bottomrule
\end{tabularx}
\end{table*}

\subsection{5.9 Multi-turn agent continuations with actual
plans}\label{multi-turn-agent-continuations-with-actual-plans}

The separately frozen v2 study completed all 92 task/policy arms: 23
SWE-bench Lite development tasks from six repositories under each of
four policies. Every arm executed at least one action and received one
nonempty task-specific Planner response when its first eligible request
was held and discarded. The agent then selected subsequent actions
rather than replaying a prescribed command. The 92 arms executed 560
actions. Full replay from the pinned base commits matched all 92 raw
records and all 92 source-minimal bundle records, including tool-result
strings or hashes, every before/after Python-source hash map, policy
eligibility, and the ordering of guidance relative to executed actions.
A separate aggregate recount matched the reported totals.

The development-cohort table reports first observed durable
Python-content changes. Six tasks from four repositories produced such a
change under at least one policy; the 23 mutation-bearing arms all
changed the withheld reference-patch Python path. Plans preceded the
first observed change in every mutation-bearing arm. This is a statement
about those observed arms, \textbf{not} a population guarantee or a
repair-success result. Seventeen arms in each of the first three
policies, and 18 all-request arms, received a plan but had no observed
Python-content change before stopping. Their plans cannot be classified
as unnecessary: the agent may have benefited from guidance during reads
or may have been unable to finish within the action cap. Planner token
and wall-time columns measure the actual local calls. The arms ran in
fixed policy order on one laptop, so differences in wall time are
descriptive and may reflect system conditions.

\begin{table*}[t]
\centering
\footnotesize
\setlength{\tabcolsep}{4pt}
\renewcommand{\arraystretch}{1.14}
\caption{Multi-turn outcomes across 23 tasks per policy.
\protect\texttt{Changed} counts arms with at least one observed
Python-content delta; \protect\texttt{Guided} uses changed arms as its
denominator. \protect\texttt{No\ change} counts arms with a plan and no
observed Python-content delta, not wasted work. Planner tokens are
reported output tokens; Planner seconds sum local call
durations.}\label{tbl:multiturn}
\begin{tabularx}{\textwidth}{@{}>{\hsize=2.2000\hsize\raggedright\arraybackslash}X>{\hsize=0.8000\hsize\raggedleft\arraybackslash}X>{\hsize=0.8000\hsize\raggedleft\arraybackslash}X>{\hsize=0.8000\hsize\raggedleft\arraybackslash}X>{\hsize=0.8000\hsize\raggedleft\arraybackslash}X>{\hsize=0.8000\hsize\raggedleft\arraybackslash}X>{\hsize=0.8000\hsize\raggedleft\arraybackslash}X@{}}

\toprule
Policy & Plans & Changed & Guided & No change & Tokens & Seconds \\
\midrule

Editor & 23 & 6 & 6/6 & 17 & 3,479 & 291.7 \\
Editor + literal & 23 & 6 & 6/6 & 17 & 3,479 & 207.7 \\
Capability & 23 & 6 & 6/6 & 17 & 3,479 & 207.8 \\
All requests & 23 & 5 & 5/5 & 18 & 3,446 & 205.4 \\
\bottomrule
\end{tabularx}
\end{table*}

The three selective policies held the same initial editor request in 22
tasks and a later editor request in the remaining task. In that
remaining \texttt{astroid-1866} task, the all-request gate held the
first \texttt{search\_text} request, while the other gates allowed three
read/search actions and held a later \texttt{replace\_text} request. The
all-request continuation used its eight-action budget on searches and
showed no Python delta; the other three continuations changed
\texttt{astroid/brain/brain\_builtin\_inference.py} at executed-action
index five. This is a concrete timing and behavior contrast, but one
continuation per policy and a locally generated plan cannot establish
that early guidance caused the different final state. Across all arms,
every first observed delta followed \texttt{replace\_text}; the live
cohort therefore does not test an actual shell-write bypass. The
public-trace witness audit in Section 5.8 supplies complementary,
offline shell-command eligibility evidence, not a live intervention
effect.

The prespecified feasibility criterion was met: all 23 tasks have
replay-valid records in every policy, and 23 mutation-bearing arms span
four repositories. These arms come from only six distinct tasks. Leaving
one repository out preserves guidance-before-first-delta in every
remaining mutation-bearing arm, but can leave as few as four such arms
under the first three policies and three under all-request. The finite
development cohort, one seed per task/policy, six repository clusters,
and no official tests preclude a credible general-population confidence
interval or task-resolution comparison. Fifty-six arms stopped at the
eight-action limit, 24 finished, and 12 stopped on model-response
timeouts or a repair timeout. All are retained in the denominators.

\subsection{5.10 Adaptive repository-stratified
follow-up}\label{adaptive-repository-stratified-follow-up}

The separately frozen follow-up completed all 240 planned arms: 30
disjoint test tasks from five repositories, four policies, and two fixed
seeds. It logged 1,692 executed actions. The follow-up policy table
distinguishes an eligible held request that reached the Planner
(\texttt{Attempted}) from actual plan insertion (\texttt{Delivered}).
Eight Planner calls timed out before delivery. Every observed first
regular-\texttt{.py} content delta followed a delivered plan, but there
were only 13 mutation-bearing arms, from three tasks in two
repositories. All first deltas followed \texttt{replace\_text}; this
live study still does not exercise a shell-write bypass. A delivered
plan on an arm without an observed Python delta is a descriptive event,
not proven wasted guidance. Planner output tokens and seconds count
successful calls only and exclude unreturned work in failed calls.

\begin{table*}[t]
\centering
\footnotesize
\setlength{\tabcolsep}{4pt}
\renewcommand{\arraystretch}{1.14}
\caption{Adaptive follow-up outcomes across 60 arms per policy.
\protect\texttt{Editor} intercepts structured editors;
\protect\texttt{Literal} adds command markers;
\protect\texttt{Capability} adds any shell request; \protect\texttt{All}
intercepts every non-finish request. \protect\texttt{Changed} counts any
observed Python-content delta; \protect\texttt{Guided} counts changed
arms with a plan before the first delta. \protect\texttt{Plan/no\ delta}
does not imply unnecessary delivery. The final column is reported
Planner output tokens and successful-call wall seconds, not total agent
cost.}\label{tbl:followup-policy}
\begin{tabularx}{\textwidth}{@{}>{\hsize=2.2000\hsize\raggedright\arraybackslash}X>{\hsize=0.8000\hsize\raggedleft\arraybackslash}X>{\hsize=0.8000\hsize\raggedleft\arraybackslash}X>{\hsize=0.8000\hsize\raggedleft\arraybackslash}X>{\hsize=0.8000\hsize\raggedleft\arraybackslash}X>{\hsize=0.8000\hsize\raggedleft\arraybackslash}X>{\hsize=0.8000\hsize\raggedleft\arraybackslash}X@{}}

\toprule
Policy & Attempted & Delivered & Changed & Guided & Plan/no delta & Tokens / seconds \\
\midrule

Editor & 8 & 7 & 3 & 3/3 & 4 & 1,540 / 122.4 \\
Literal & 8 & 6 & 4 & 4/4 & 2 & 1,320 / 114.4 \\
Capability & 10 & 8 & 4 & 4/4 & 4 & 1,528 / 113.4 \\
All & 58 & 55 & 2 & 2/2 & 53 & 8,980 / 794.5 \\
\bottomrule
\end{tabularx}
\end{table*}

\begin{figure*}[t]
\centering
\pandocbounded{\includegraphics[keepaspectratio,alt={The 30-task
follow-up by gate, with 60 arms per policy. Bars show delivered plans;
dots show arms with an observed Python-source change. The two quantities
have the same arm denominator but are not complements or measures of
repair success. OpenAI Codex assisted with the rendering code and
layout; the author checked the figure against the recorded
table.}]{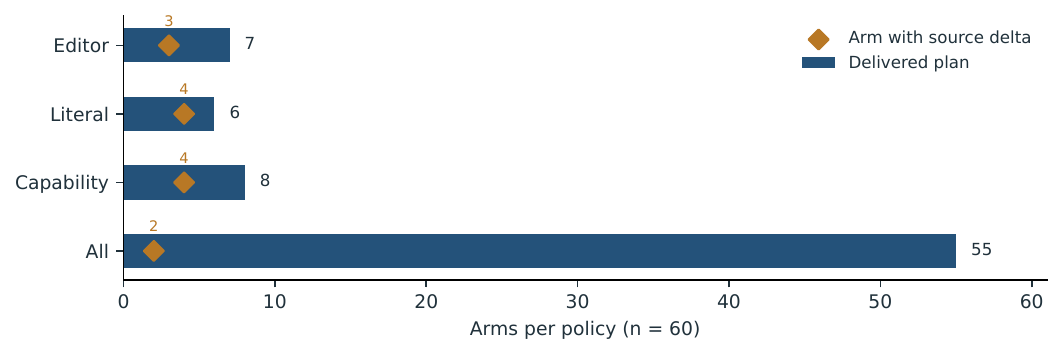}}
\caption{The 30-task follow-up by gate, with 60 arms per policy. Bars
show delivered plans; dots show arms with an observed Python-source
change. The two quantities have the same arm denominator but are not
complements or measures of repair success. OpenAI Codex assisted with
the rendering code and layout; the author checked the figure against the
recorded table.}\label{fig:followup-delivery}
\end{figure*}

The repository table shows the concentration that the aggregate 13/13
guided count can hide. Only \texttt{psf\_\_requests-2148},
\texttt{psf\_\_requests-2317}, and \texttt{pytest-dev\_\_pytest-8365}
changed source under any policy. Six of the Requests \texttt{2148} arms
changed \texttt{requests/adapters.py}, whereas that issue's withheld
reference patch names \texttt{requests/models.py}. The remaining seven
mutation-bearing arms ended with a change on their task's withheld
reference-patch Python path. Such overlap is a relevance check, not a
correctness or test-pass result. All five pytest \texttt{8365} changes
insert a call to \texttt{sanitize\_username}, a name absent from the
pinned base commit and not defined by later recorded actions; the
reference patch instead catches \texttt{OSError} around directory
creation. Reconstructing the five final files from the recorded edits
reproduces their hashes, and an isolated call to each edited
\texttt{get\_user()} raises \texttt{NameError}. Omitting Requests leaves
five mutation-bearing arms; omitting pytest leaves eight. The
predeclared feasibility condition \textbf{failed}: complete records
exist for all 60 task/seed pairs and four policies, but only 13 changed
arms in two repositories fell short of 20 in three. We therefore do not
attach a task-level binomial interval or a general delivery rate to the
conditional 13/13.

Post hoc local tests further separate a source change from a repair. For
each of the three changed tasks, we applied the pinned test patch to
fresh base, reference-patch, and reconstructed agent workspaces. A
selected FAIL\_TO\_PASS test fails on the base and passes with the
reference patch in every task. It still fails for all six Requests
\texttt{2148} arms and all five pytest \texttt{8365} arms. Both Requests
\texttt{2317} arms pass one selected target using a minimal local HTTP
fixture. On pytest \texttt{8365}, all 32 PASS\_TO\_PASS selectors pass
on the base and reference workspaces; the agent edit yields one failure
and 31 setup errors. These are controlled local diagnostics, not
official SWE-bench grades. We did not run the remaining Requests tests,
so neither the two target passes nor any gate comparison establishes
repair success.

\begin{table*}[t]
\centering
\footnotesize
\setlength{\tabcolsep}{4pt}
\renewcommand{\arraystretch}{1.14}
\caption{Repository concentration in the 30-task follow-up.
\protect\texttt{Reference\ path} counts changed arms whose final state
also differs on the withheld reference-patch Python path; no
issue-resolution grading was
run.}\label{tbl:followup-repos}
\begin{tabularx}{\textwidth}{@{}>{\hsize=2.2000\hsize\raggedright\arraybackslash}X>{\hsize=0.7600\hsize\raggedleft\arraybackslash}X>{\hsize=0.7600\hsize\raggedleft\arraybackslash}X>{\hsize=0.7600\hsize\raggedleft\arraybackslash}X>{\hsize=0.7600\hsize\raggedleft\arraybackslash}X>{\hsize=0.7600\hsize\raggedleft\arraybackslash}X@{}}

\toprule
Repository & Tasks & Arms & Changed & Reference path & Delivered plans \\
\midrule

Requests & 6 & 48 & 8 & 2 & 16 \\
Xarray & 5 & 40 & 0 & 0 & 10 \\
Pylint & 6 & 48 & 0 & 0 & 17 \\
pytest & 7 & 56 & 5 & 5 & 21 \\
Sphinx & 6 & 48 & 0 & 0 & 12 \\
\textbf{Total} & \textbf{30} & \textbf{240} & \textbf{13} & \textbf{7} &
\textbf{76} \\
\bottomrule
\end{tabularx}
\end{table*}

\begin{figure*}[t]
\centering
\pandocbounded{\includegraphics[keepaspectratio,alt={Source changes in
the follow-up were concentrated in Requests and pytest. Other three
repositories had no observed Python-content delta within the
eight-action limit. The shaded segment shows final changes on a withheld
reference-patch Python path; path overlap is not task correctness.
OpenAI Codex assisted with the rendering code and layout; the author
checked the figure against the recorded
table.}]{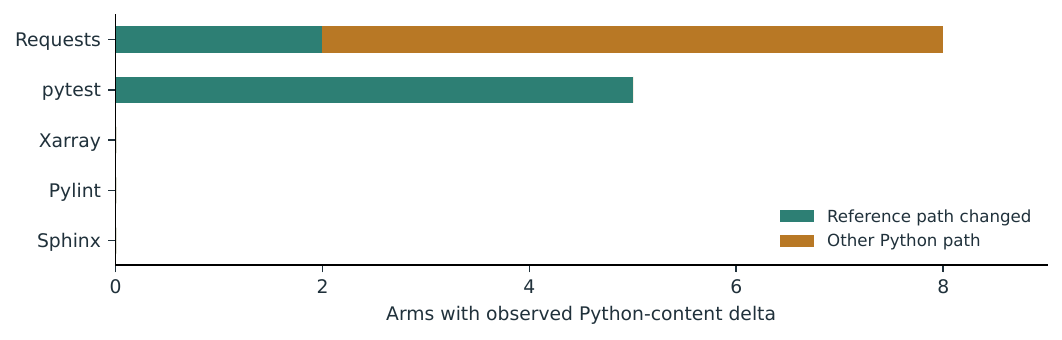}}
\caption{Source changes in the follow-up were concentrated in Requests
and pytest. Other three repositories had no observed Python-content
delta within the eight-action limit. The shaded segment shows final
changes on a withheld reference-patch Python path; path overlap is not
task correctness. OpenAI Codex assisted with the rendering code and
layout; the author checked the figure against the recorded
table.}\label{fig:followup-repos}
\end{figure*}

The policy contrasts run in both directions. On Requests \texttt{2317},
all-request held an initial \texttt{read\_file}, delivered a plan, and
the agent immediately chose a \texttt{replace\_text} that changed the
reference-patch file in both seeds; the narrower gates made no source
change on that task. On pytest \texttt{8365}, all-request held an
initial \texttt{search\_text} and then used eight actions reading
without a change in either seed. The selective gates instead allowed
search and read actions, held a later \texttt{replace\_text}, and five
of six arms changed the reference-patch file; the remaining editor-only
arm timed out before changing it. These are paired continuations under
different policies, not proof that one timing produces better repairs.
On pytest \texttt{5221}, capability and all-request held a proposed
inspection command, \texttt{pytest\ -\/-fixtures}, while editor-only and
literal-marker gates executed it. A general shell's write capability
thus produced an actual plan-delivery cost even when the proposed
command itself was read-oriented; none of those arms changed source.

The advisory inspection-first request yielded an accepted initial
list/read/search in 208/240 arms. Eight initially proposed a shell
command, 22 had an invalid first response, and two timed out before any
response. Of 120 task/policy seed pairs, both first model responses were
available in 118 and were identical in all 118. Executed action
sequences matched in 108 pairs, stop reasons in 112, and mutation status
in 119. With temperature zero, two seed values supplied little
behavioral diversity; timeouts explain much of the variation. Across
policies, 181 arms reached the eight-action limit, 24 finished, 19 had
model-response timeouts, eight had Planner timeouts, and eight failed
format repair. All remain in the denominators. One pytest task
repeatedly led this local agent to request \texttt{pmxbot/} files
mentioned in an external reproducer, which were absent from the pytest
checkout. This illustrates the model/navigation limit of the observed
mutation rate. The follow-up is an adaptive robustness check, not an
independent confirmation of the development cohort or a repair-effect
experiment.

\subsection{5.11 A pilot-guided 14B
cohort}\label{a-pilot-guided-14b-cohort}

The final, separately frozen cohort completed all 76 arms on 19 tasks.
Its 381 executed actions include 50 requests held for planning: 48 plans
were delivered and two Planner calls timed out. Fifteen arms changed the
content of a regular Python file, across seven tasks in three
repositories. Guidance preceded every observed first change. The
editor-only gate delivered 12 plans in 38 arms and preceded all five
changes under that policy; the all-request gate delivered 36 plans in 38
arms and preceded all ten changes under that policy. All 15 first
changes followed \texttt{replace\_text}. The live runs therefore still
do not exercise a shell-first source change, despite the shell-write
cases in the controlled suite and public traces. Twelve changed arms
ended with a difference on the withheld reference-patch Python path;
this is a path-overlap proxy, not repair grading.

\begin{figure*}[t]
\centering
\pandocbounded{\includegraphics[keepaspectratio,alt={Python-source
changes by task in the 76-arm, 14B cohort. Each cell counts the two
repeats of one task under one policy. The color records a content delta,
not a correct repair. Task and repository clustering is visible; the
four repositories are Seaborn, Flask, pytest, and Sphinx. OpenAI Codex
assisted with the rendering code and layout; the author checked the
figure against the recorded
summary.}]{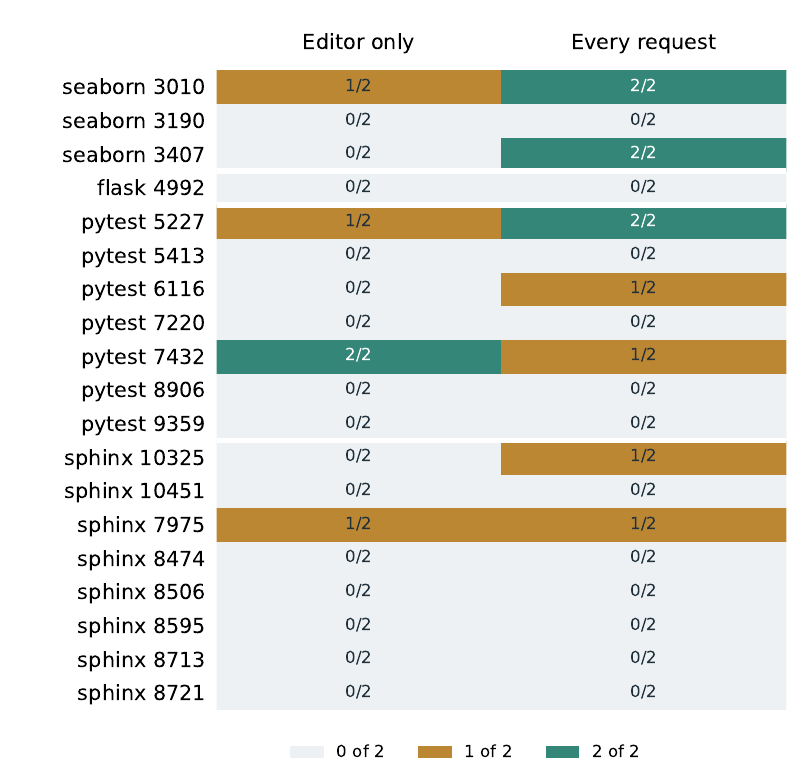}}
\caption{Python-source changes by task in the 76-arm, 14B cohort. Each
cell counts the two repeats of one task under one policy. The color
records a content delta, not a correct repair. Task and repository
clustering is visible; the four repositories are Seaborn, Flask, pytest,
and Sphinx. OpenAI Codex assisted with the rendering code and layout;
the author checked the figure against the recorded
summary.}\label{fig:small-patch-tasks}
\end{figure*}

The figure shows why the aggregate 15/76 should not be treated as 15
independent tasks. Seven tasks supplied all changes; Flask supplied
none. The prespecified feasibility target required at least 20 changing
arms in at least three repositories. The repository part was met, but
the arm-count part \textbf{failed}. A task-cluster bootstrap within the
four fixed repositories gives a descriptive all-request minus
editor-only change-rate difference of 0.132, with a 95\% percentile
interval of 0.026--0.263. This interval is conditional on the selected
tasks and policies; Flask has only one task and contributes no
within-repository sampling variation. It does not establish a general or
causal advantage for all-request guidance.

The temperature-0.7 repeats varied more than the earlier
zero-temperature repeats. Within the same task and policy, 12/38 repeat
pairs began with identical model text and only 1/38 executed the same
action sequence. All 38 paired policy runs began with identical first
model text under their shared seed, before either gate could intervene;
none had an identical full executed sequence. This is evidence of varied
continuations in this local setup, not 76 independent strategies.
Sixteen arms ended on model timeouts, two on Planner timeouts, three at
the action limit, and 55 with \texttt{finish}; all remain in the
denominator.

Source change also proved a weak repair proxy. Replaying all 15 changed
arms reproduced their final Python-file hashes. Four final changed arms
have at least one edited Python file that fails \texttt{compile}, and
four arms replace a file with less than half its original line count;
these groups overlap. One pytest arm shrinks a 662-line module to eight
lines yet passes the syntax check. We have not run the official
SWE-bench grader on this cohort, so no changed arm is classified as a
resolved issue.

We then ran a separately frozen, \emph{post hoc} local test diagnostic
on all 15 changed final Python-only patches. For each of seven tasks,
the unmodified base failed at least one listed \texttt{FAIL\_TO\_PASS}
test, the reference patch passed all listed targets, and both controls
passed the usable regression selectors in the same local environment.
Seven changed arms passed their local targets and tested regressions;
five failed a target; one passed its target but failed a regression; two
could not collect because the edited module was invalid. These are arm
counts from 13 distinct final patches, not independent repair successes.
Six arms in two pytest tasks used a \textbf{partial regression subset}:
the released \texttt{PASS\_TO\_PASS} fields contain eight malformed or
transcript-fragment selectors in total. The omission rule was recorded
before testing either task's agent patches, and all original IDs remain
in the result records. Three of the seven local target-and-regression
passes used full listed regression sets; four used partial sets. The
other 61 arms had no final observed Python change and were not graded.
The diagnostic neither reproduces the official environment nor
establishes a gate effect on repair quality.

\begin{table*}[t]
\centering
\footnotesize
\setlength{\tabcolsep}{4pt}
\renewcommand{\arraystretch}{1.14}
\caption{Post hoc local target and regression checks on the 15 changed
14B arms. \protect\texttt{Pass} requires listed targets and tested
regressions to pass under valid base/reference controls.
\protect\texttt{Target\ fail} includes arms whose target failed, whether
or not a regression also failed. \protect\texttt{Error} denotes
collection or setup errors from an invalid edited module. Six pytest
arms use partial regression lists; all other arms use full listed sets.
No count is an official SWE-bench
resolution.}\label{tbl:small-patch-local-outcomes}
\begin{tabularx}{\textwidth}{@{}>{\hsize=2.2000\hsize\raggedright\arraybackslash}X>{\hsize=0.7600\hsize\raggedleft\arraybackslash}X>{\hsize=0.7600\hsize\raggedleft\arraybackslash}X>{\hsize=0.7600\hsize\raggedleft\arraybackslash}X>{\hsize=0.7600\hsize\raggedleft\arraybackslash}X>{\hsize=0.7600\hsize\raggedleft\arraybackslash}X@{}}

\toprule
Repository & Changed & Pass & Target fail & Target pass, regression fail & Error \\
\midrule

Seaborn & 5 & 3 & 2 & 0 & 0 \\
pytest & 7 & 4 & 1 & 1 & 1 \\
Sphinx & 3 & 0 & 2 & 0 & 1 \\
\textbf{Total} & \textbf{15} & \textbf{7} & \textbf{5} & \textbf{1} &
\textbf{2} \\
\bottomrule
\end{tabularx}
\end{table*}

The raw action-state chain verifies all 76 arms. Independent execution
of the recorded actions reproduces exact tool outputs for 75; one
Seaborn arm differs only in an ephemeral checkout prefix printed in
three \texttt{unittest} tracebacks. A separately retained strict report
records that byte-output mismatch. Under the explicitly scoped prefix
normalization, its action order, exit codes, outputs, and every
source-state map match. Replay does not authenticate the original
model's unshared conversation text.

\section{6. What These Results Mean}\label{what-these-results-mean}

The phrase ``guidance before the first edit'' hides five questions.
Which policy was assigned? Which requests did its rule cover? Did the
gate fire? Did a plan reach the Executor? When did source bytes first
differ? An agent can change a file through an action outside the gate,
so these events do not always form a neat funnel. A stored plan also
tells us that text was placed in a trace, not how the model used it.

The constructed cells show the trade-off plainly. Editor-only guidance
reaches 3 of 24 chosen package-source changes; an all-request gate
reaches all 24 and also delivers guidance on all 15 actions without an
immediate package-source change. That second count measures delivery
frequency in this suite. It does not say those plans would be wasted
during a longer repair. PAUSE and SI2CA show why public traces alone
cannot give the same before/after comparison: they record commands and
final patches, but not synchronized source bytes at every step.

Actual agent continuations add a harder test because a delivered plan
can change the next request. In the development cohort, however, 22 of
23 tasks first reached an editor request under every policy. Those tasks
gave the four gates little opportunity to behave differently. In public
shell traces, the capability gate would recognize all 19 selected writer
commands, but it would also hold their 38 flanking read slots: the same
shell tool permits both. This is a property of the tool interface and
the selected witnesses, not a measured field error rate.

The follow-up finally separates the gates in live runs: selective
policies deliver 6--8 plans per 60 arms, while all-request delivers 55.
Eight held requests reach the Planner but never yield a plan. Two
\texttt{pytest\ -\/-fixtures} inspections are held by the capability
gate because they use a shell that \emph{could} write. Guidance precedes
all 13 observed first source changes, but those changes arise in only
three tasks and two repositories. The study therefore fails its own
coverage threshold. Requests and pytest also show opposite changes in
behavior after early guidance. Without reliable repair outcomes and
diverse repeats, we cannot call either timing strategy better for fixing
issues.

The 14B cohort changes that picture in one respect: only one of 38
within-policy repeat pairs follows the same complete action sequence. It
still misses the 20-changing-arm target, and its 15 first source changes
all use \texttt{replace\_text}. More varied action sequences therefore
do not, by themselves, validate shell-write coverage or produce enough
sound patches. Four changed arms fail Python syntax checks, and a
syntactically valid eight-line replacement of a 662-line module shows
why even a passing compile check is a thin outcome measure. The post hoc
local tests found target-and-regression passes in only seven of 15
changed arms, with partial regression coverage in two of those passes.
The all-request gate produces more source-changing arms in this selected
cohort, but the same local model generates both plans and actions, and
the policies alter entire continuations after delivery. The local check
does not provide an official repair grade, so the observed difference is
neither a correctness gain nor a general gate ranking.

\subsection{6.1 A falsifiable delivery
contract}\label{a-falsifiable-delivery-contract}

A reusable experiment should log an intervention policy version,
configured source-path set, task and run IDs, ordered tool-call IDs,
proposed action, eligibility decision, delivery event, and execution
result. To audit the stronger promise ``before any source mutation,'' it
should also record the relevant workspace state immediately before and
after each action, including tracked, staged, and newly created files; a
plain \texttt{git\ diff} misses some of these. For large workspaces, a
watcher can nominate changed paths, but a content hash or equivalent
state check should confirm the delta. Tool-call and workspace events
need one total order so an inserted plan can be placed before the first
confirmed mutation. This state stream can distinguish a failed call with
no change, a failed call that partially changed files, a successful
no-op, and a real mutation.

The intervention must also state which guarantee it implements. A hook
that sees only structured editor requests promises a plan before the
first \emph{eligible editor request}, not before all repository writes.
A stronger pre-mutation guarantee requires gating every writable tool
surface, restricting the shell until guidance is delivered, or enforcing
a write barrier outside the tools. Merely scanning command text cannot
certify that guarantee: shell commands can execute scripts, copies,
pipelines, and dynamically assembled paths. The controlled suite
demonstrates marker misses and false positives on selected forms, not a
field failure rate for command scanning.

An extended instrumented validation suite should include symlinks,
transient writes reverted within an action, changes outside the observed
root, and concurrent writers in addition to the forms tested here. It
should also exercise actual PAUSE hooks with an authenticated path
configuration and compare delivery to observed state. The 39-cell suite
validates its observer against target bytes; the multi-turn study
validates plan delivery in a constrained local agent. Neither tests the
historical PAUSE implementation against atomic workspace state.

The earlier twelve-issue local-agent pilot had five runs fail at the
first JSON action and only one mutation-bearing run; it motivated the
separately frozen development protocol but is not pooled with either
multi-turn cohort. Both later studies observe durable \texttt{.py}
content across each executed action in a write-confined temporary
repository. This misses transient in-action writes later reverted,
symlinks, non-Python source, and any unlogged concurrent process. A
production pre-mutation guarantee would require a write barrier or
filesystem event stream over every writable surface, with state checks
to verify the recorded effects.

\subsection{6.2 Threats to validity}\label{threats-to-validity}

\textbf{Construct validity.} A final diff of a pre-existing \texttt{.py}
file is a narrow proxy for source modification. It omits other
languages, intermediate changes later reverted, and modifications hidden
by truncated or missing patches. A command naming that path may write
nothing or write a different file. Conversely, a helper can assemble its
target dynamically and evade literal matching. The inactive PAUSE TeX
prefix list fits 465 SWE-agent and 260 Pi positive trigger paths and 36
selected direct-editor exclusions but is not authenticated as the live
harness configuration. Multiple rules could fit these observations, so
configured eligibility cannot be established case by case.

The ten SI2CA adjacent-read cases reduce the command-versus-delta
ambiguity for a strict subset, but the identical-read criterion has
unknown recall and the preliminary text-difference review was not
independently human-coded. Each comparison brackets a command in the
logged action sequence rather than observing an atomic filesystem
transition; concurrent or unlogged processes remain a theoretical
alternative.

The seven PAUSE Pi adjacent-read runs have the same observation limit.
They are concentrated in two repositories, so they should not be treated
as seven independent demonstrations of the frequency of missed triggers
in other repository layouts. The actual configured package-directory
list is still needed to decide whether a particular path met the editor
hook's path predicate; a shell change can be outside that narrower
contract regardless.

The triggered SymPy pre-plan case was selected after a four-run
candidate audit. The overlapping-read detector was then designed with
knowledge of that case; it has unknown recall because most triggered
runs lack comparable bounded reads. Its four differing pairs in one run
therefore supply neither a population mutation rate nor independent
discovery evidence. The case establishes a changed anchored function
region before the recorded plan message, subject to the same
unlogged-writer caveat, but supplies no exact first-change timestamp.
Subsequent commands in that trace include a checkout, so the pre-plan
change should not be equated with the final patch.

\textbf{Internal validity.} The publisher datasets do not expose
synchronized pre/post workspace snapshots for each tool call. We
therefore cannot identify the first actual mutation in those releases or
claim that a particular candidate caused the final diff. Task-matched
PAUSE arms are independent trajectories, and delivery is
post-assignment. SI2CA standard and discovered arms also generate
different trajectories; their candidate rates are descriptive and must
not be compared as a causal effect of the gate. In our local studies,
policy arms share a task and initial source state, but the development
cohort has one continuation each and the 7B follow-up's two
temperature-zero seeds produce largely identical trajectories. The 14B
repeats vary more, but run order, runtime variance, timeouts, and
task-specific Planner text still prevent attributing a final file or
latency difference solely to the gate. The development cohort used fixed
policy order; later cohorts rotated it but did not remove hardware or
clock variation. The raw logs are retained locally; a source-minimal
replay bundle can verify actions and source states but cannot
authenticate the model's exact generated text or plan wording without
those logs.

\textbf{External validity.} PAUSE and SI2CA use different interventions
and task pools; the latter validates lexical and gate-phase concerns
rather than replicating PAUSE's delivery rate. The PAUSE Python-path
audit does not cover non-Python source or private production
repositories. The two manual samples were selected deterministically
after exploratory detector development, inspected in one preliminary
pass without independent human adjudication, and cannot establish
detector recall or inter-rater reliability. The development cohort
covers all 23 tasks in six repositories, but only six tasks in four
repositories changed observed source. The adaptive 7B test-split
follow-up includes 30 tasks from five other repositories, yet only three
tasks in two repositories changed source; six of its 13 changed arms
edited a different file from the withheld reference-patch Python path.
Its near-deterministic repeats, 90-second limit, and absence of complete
task grading limit transfer. The later 14B cohort selects small
reference patches from four available repositories and uses a model
chosen after separate pilots; it is not a random held-out sample of
coding tasks. Its 300-second limit and structured action schema still
constrain behavior, and its 15 changes lack official grading. Across
both later cohorts, all first changes used a structured editor, so live
shell-bypass prevalence remains unmeasured. The local target tests in
the 7B cohort cover only the three changed tasks and were selected after
seeing which tasks changed source. The post hoc strict witness intervals
are selected positives and cannot estimate field precision or recall for
the four owned gates. Counts characterize the pinned releases, local
model configurations, and source commits, not all coding agents or
future versions.

The exact historical PAUSE path list is no longer a prerequisite for the
owned-gate results; it remains a limit on case-level interpretation of
the secondary PAUSE analysis. The 7B follow-up misses its
mutation-coverage and behavioral-repeat targets. The 14B cohort improves
repeat diversity and reaches three repositories, but also misses the
20-arm target and yields several invalid or severely shortened files. A
full EMSE submission still needs outcome-aware evidence, independent
human coding and scientific review, and a public reproduction package
with rights-cleared inputs. The local studies show when guidance reached
constrained agents, not whether planning improved repair quality.

\section{7. Conclusion}\label{conclusion}

An intervention is only as early as the requests its gate can see and
the plans it actually delivers. Our action-level source checks place
every observed first Python-content change after guidance in the
recorded local cohorts. That result needs a narrow reading. The 7B
follow-up changes source in only three tasks, while the more varied 14B
cohort changes source in seven tasks and still misses its preset
coverage target. Every live first change uses a structured editor;
several changed final files are invalid or severely shortened. Public
PAUSE and SI2CA traces add changed-text witnesses but cannot recover
exact first-mutation times. The audit makes delivery timing testable. It
does not yet establish a repair benefit or general pre-mutation
coverage. Complete outcome testing, independent witness review, and a
rights-cleared public artifact remain necessary for a journal
submission.

\section{Statements and Declarations}\label{statements-and-declarations}

\textbf{Funding.} This research was self-funded by the author.

\textbf{Competing interests.} The author declares no relevant financial
or nonfinancial competing interests.

\textbf{Ethics and consent.} This study analyzed public software-agent
trajectories and locally executed software agents. No human participants
were recruited or experimentally intervened upon. Consent to participate
was therefore not applicable. No institutional ethics approval or
exemption determination is claimed.

\textbf{Data and code availability.} The public PAUSE and SI2CA sources
and pinned revisions are cited in the replication guide. This project's
repository and full local model transcripts remain private. No public
replication archive has been approved.

\textbf{AI Declaration.} The author specified the figures' content and
design. OpenAI Codex assisted in converting those designs into the eight
code-rendered figures, including plotting code and layout. The author
checked every figure against its underlying data or study protocol. The
scripts and inputs are identified in the research repository; the exact
Codex model version used during figure preparation was not retained.
OpenAI ChatGPT and Codex also supported the research and writing tasks
described in Section 4.3; preliminary trace checks were tool-assisted,
not independent human coding. The author remains responsible for the
manuscript and figures.

\section{References}\label{references}

\begin{itemize}
\tightlist
\item
  AgentLens contributors (2026).
  \href{https://github.com/dreadnode/agent-lens/tree/9eab2b0ca729947cbfed3ecd8cab3f45521447cf}{AgentLens:
  multi-session coding-agent trajectory harness and shadow-git state
  tracking}. Open-source implementation; pinned revision in the
  replication guide.
\item
  Fan, R.-Z., Zhang, Z., Ma, S., Hu, Y., Wang, S., Song, K., Liu, F.,
  Zamani, H., and Wang, X. (2026).
  \href{https://arxiv.org/abs/2609.20804}{An Empirical Study of Harness
  Design for Coding Agents}. arXiv:2609.20804.
\item
  He, J., Shi, M., Jia, J., Yang, X., and Sun, D. (2026).
  \href{https://arxiv.org/abs/2608.22960}{What Process Evaluation of
  Coding Agents Actually Measures: Action, Task, and Step Are Three
  Different Levels}. arXiv:2608.22960.
\item
  Jimenez, C. E., Yang, J., Wettig, A., Yao, S., Pei, K., Press, O., and
  Narasimhan, K. R. (2024).
  \href{https://proceedings.iclr.cc/paper_files/paper/2024/hash/edac78c3e300629acfe6cbe9ca88fb84-Abstract-Conference.html}{SWE-bench:
  Can Language Models Resolve Real-world Github Issues?}. ICLR 2024.
  \href{https://huggingface.co/datasets/princeton-nlp/SWE-bench_Lite/tree/6ec7bb89b9342f664a54a6e0a6ea6501d3437cc2/data}{Lite
  development dataset}.
\item
  Kapner, B., Soceanu, C., Petrunin, A., and Gartner, H. (2026).
  \href{https://arxiv.org/abs/2609.07360v3}{Scanning the Harness:
  Configuration Exposures in AI Coding-Agent Supply Chains}.
  arXiv:2609.07360v3.
\item
  Mirzoian, V. (2026).
  \href{https://github.com/Vlad-Mirzoian/retrace}{Retrace}. Project
  documentation and implementation, inspected 24 October 2026.
\item
  Nian, Y., Yuan, A., Zhang, H., Li, J., Li, L., Hu, X., Wei, H., Xiao,
  X., Xiao, C., and Zhao, Y. (2026).
  \href{https://arxiv.org/abs/2604.05485}{Auditable Agents}.
  arXiv:2604.05485, version 4.
\item
  Oderinwale, H. (2026). \href{https://arxiv.org/abs/2606.16988}{Agent
  trajectories as programs: fingerprinting and programming coding-agent
  behavior}. arXiv:2606.16988.
  \href{https://github.com/hamidahoderinwale/procgrep}{ProcGrep
  implementation}.
\item
  Patel, V., Tan, B., Xue, J., and Sun, Y. (2026).
  \href{https://arxiv.org/abs/2610.29870v1}{PAUSE: Search-Free Steering
  of Coding Agents Before Their First Source-Code Edit}.
  arXiv:2610.29870v1.
  \href{https://huggingface.co/datasets/pause-fse27-artifacts/PAUSE-trajectories}{Trajectory
  dataset}.
\item
  SI2CA Team (2026).
  \href{https://github.com/Self-Improving-Coding-Agents/SI2CA/blob/main/reports/SI2CA_Technical_Report.pdf}{(Self-Improving)²
  Coding Agents: Curating High-Quality Trajectories via Recursive
  Self-Improvement}. Technical report.
  \href{https://huggingface.co/datasets/Self-Improving-Coding-Agents/SI2CA-Training-Trajectories}{Trajectory
  dataset};
  \href{https://github.com/Self-Improving-Coding-Agents/SI2CA/blob/e750e3512e4368e4e6a3d3a7392a73dc50bad12a/si2ca/runtime/protocol.py}{source
  implementation}.
\item
  Trace Commons (2026).
  \href{https://huggingface.co/datasets/trace-commons/agent-traces/tree/112ebd4d03ce852b00e935d523107c3d0c9a65bf}{Agent
  traces}. Public dataset; pinned revision in the replication guide.
\item
  Xie, Y., Zhang, G., Wang, X., Roth, D., Fried, D., and Rosé, C.
  (2026).
  \href{https://catradar-coding-agent.github.io/static/pdfs/CATRadar.pdf}{Successful
  Rollouts are Not Equal: Discovering Influential Behaviors for Training
  Coding Agents}. Author preprint.
\item
  Xiao, S., Wang, S., Chen, X., Chao, K., Cui, M., Qian, F., Mei, C.,
  Meng, F., Yu, Z., and Yin, J. (2026).
  \href{https://arxiv.org/abs/2610.00372}{When Harnesses Lose the
  Signal: Causal Evaluation of Recovery in LLM Agents}.
  arXiv:2610.00372.
\item
  Yang, B., Li, Z., Yang, Z., Jia, K., Yin, X., Liu, M., and Tian, H.
  (2026). \href{https://arxiv.org/abs/2610.04375}{Do Tool Calls Execute
  as Intended? Measuring and Repairing Intent-Execution Correspondence
  in LLM Agents}. arXiv:2610.04375.
\item
  Zhong, H., and Zhu, S. (2026).
  \href{https://arxiv.org/abs/2605.13357}{AI Harness Engineering: A
  Runtime Substrate for Foundation-Model Software Agents}.
  arXiv:2605.13357.
\item
  Zhu, Z., Huang, L., Ji, R., Ye, M., Zhou, Z., Guo, H., Wu, J., Ye, Y.,
  Feng, Y., Dai, H.-N., and Zheng, Z. (2026).
  \href{https://arxiv.org/abs/2610.07639v2}{HarnessSecurity-Bench: Do
  Security Mechanisms Really Protect Coding Agent Harnesses?}.
  arXiv:2610.07639v2.
\end{itemize}

\section{Appendix A. Executable intervention
rule}\label{appendix-a.-executable-intervention-rule}

The local harness evaluates a proposed action before executing it. The
simplified code below preserves the implemented decision order in
\texttt{scripts/run\_multiturn\_agent\_study.py};
\texttt{MUTATION\_MARKERS} is the fixed nine-string tuple declared in
the runner. \texttt{finish} is never eligible. The Planner is called at
most once in an arm.

\begin{Shaded}
\begin{Highlighting}[]
\KeywordTok{def}\NormalTok{ eligible(policy, action):}
\NormalTok{    name }\OperatorTok{=}\NormalTok{ action[}\StringTok{"name"}\NormalTok{]}
    \ControlFlowTok{if}\NormalTok{ name }\OperatorTok{==} \StringTok{"finish"}\NormalTok{:}
        \ControlFlowTok{return} \VariableTok{False}
    \ControlFlowTok{if}\NormalTok{ policy }\OperatorTok{==} \StringTok{"all\_requests"}\NormalTok{:}
        \ControlFlowTok{return} \VariableTok{True}
    \ControlFlowTok{if}\NormalTok{ name }\KeywordTok{in}\NormalTok{ \{}\StringTok{"replace\_text"}\NormalTok{, }\StringTok{"write\_file"}\NormalTok{\}:}
        \ControlFlowTok{return} \VariableTok{True}
    \ControlFlowTok{if}\NormalTok{ name }\OperatorTok{!=} \StringTok{"shell"}\NormalTok{:}
        \ControlFlowTok{return} \VariableTok{False}
    \ControlFlowTok{if}\NormalTok{ policy }\OperatorTok{==} \StringTok{"writable\_capability"}\NormalTok{:}
        \ControlFlowTok{return} \VariableTok{True}
    \ControlFlowTok{if}\NormalTok{ policy }\OperatorTok{==} \StringTok{"editor\_plus\_literal"}\NormalTok{:}
\NormalTok{        args }\OperatorTok{=}\NormalTok{ action[}\StringTok{"arguments"}\NormalTok{]}
\NormalTok{        command }\OperatorTok{=} \BuiltInTok{str}\NormalTok{(args.get(}\StringTok{"command"}\NormalTok{, }\StringTok{""}\NormalTok{))}
        \ControlFlowTok{return} \BuiltInTok{any}\NormalTok{(m }\KeywordTok{in}\NormalTok{ command}
                   \ControlFlowTok{for}\NormalTok{ m }\KeywordTok{in}\NormalTok{ MUTATION\_MARKERS)}
    \ControlFlowTok{return} \VariableTok{False}  \CommentTok{\# editor\_only}
\end{Highlighting}
\end{Shaded}

The \emph{literal} strategy searches command text, including false
matches in harmless commands; it cannot identify a write performed by an
indirectly invoked helper when no marker appears. The \emph{capability}
strategy treats the general shell as writable even when the proposed
command only reads. These are policy definitions, not classifications
learned from the follow-up outcomes.

\section{Appendix B. State and replay
checks}\label{appendix-b.-state-and-replay-checks}

The source observer inventories regular \texttt{.py} files and hashes
their bytes. Immediately before and after each executed action, it
compares the maps. The first unequal map identifies the first
\textbf{observed durable Python-content delta} at an action boundary. It
does not see a temporary write reverted inside one action, a non-Python
file, or a symlink. Guidance must be delivered before the action whose
after-state first differs.

\begin{Shaded}
\begin{Highlighting}[]
\NormalTok{before }\OperatorTok{=}\NormalTok{ snapshot(repo)}
\NormalTok{result, exit\_code }\OperatorTok{=}\NormalTok{ execute(action)}
\NormalTok{after }\OperatorTok{=}\NormalTok{ snapshot(repo)}
\NormalTok{paths }\OperatorTok{=}\NormalTok{ before.keys() }\OperatorTok{|}\NormalTok{ after.keys()}
\NormalTok{changed }\OperatorTok{=}\NormalTok{ [p }\ControlFlowTok{for}\NormalTok{ p }\KeywordTok{in} \BuiltInTok{sorted}\NormalTok{(paths)}
           \ControlFlowTok{if}\NormalTok{ before.get(p) }\OperatorTok{!=}\NormalTok{ after.get(p)]}
\NormalTok{record(action, before, after,}
\NormalTok{       changed, result, exit\_code)}
\end{Highlighting}
\end{Shaded}

The pre-run checks pin the dataset Parquet hash, manifest, task ID, and
base commit. Each arm starts in a fresh extracted checkout. At a held
eligible request, the runner snapshots the workspace before and after
the Planner call and asserts that planning did not change it. The
replayer then rebuilds the base state and executes the recorded actions
without model calls, checking the policy decision, result, before/after
hashes, and final state. In the 240-arm follow-up, 236 raw records
reproduce exact output text. Four \texttt{pytest}/\texttt{pip}
network-denied retry outputs vary; a narrower check reproduces their
order, exit codes, source hashes, and final state. Replaying actions
cannot independently verify that the original model generated the stored
actions or plan text.

The 76-arm 14B cohort has 75 byte-exact action-output replays and one
scoped exception. In Seaborn task \texttt{3010}, replicate 1 under
all-request, three \texttt{unittest\ discover} results print the fresh
temporary checkout path inside otherwise identical import-error
tracebacks. Normalizing only that ephemeral path makes all three outputs
agree; all action order, return codes, and Python-source hash maps
already agree. The strict 75/76 report is retained alongside the scoped
76/76 result. A separate final-state diagnostic replays all 15 changed
arms and checks edited \texttt{.py} files with Python \texttt{compile};
it is not a task test or repair grade.

\section{Appendix C. Reading the evidence
sources}\label{appendix-c.-reading-the-evidence-sources}

The 39 prescribed cells test policy semantics against known target-file
bytes; they have placeholder guidance and no agent continuation. The
development cohort uses 23 real issue descriptions and one agent
continuation per policy. The 7B adaptive follow-up uses 30 disjoint test
issues and two fixed seeds per policy, but its repeats are almost
identical at temperature zero. The separate 14B small-patch cohort uses
19 selected test issues, temperature 0.7, and two repeats under each of
two policies; it was chosen after separate model pilots. PAUSE and SI2CA
are secondary trace audits with different publisher interventions and
incomplete historical file-state records. Their changed-read witnesses
identify bounded intervals of changed text, not exact mutation instants.
None of these layers grades the agent's patch with the official
SWE-bench evaluator. The sample and validation limits in Section 6 apply
to every figure in this paper.

\end{document}